\pdfoutput=1
\documentclass[article,12pt]{elsarticle}

\usepackage{amssymb}
\usepackage{amsmath}
\usepackage{float}
\usepackage{subfigure}
\usepackage{algorithm}
\usepackage{subcaption}
\usepackage{algpseudocode}
\usepackage{hyperref}

\journal{Chaos Solitons \& Fractals}

\usepackage{color}

\begin{document}
	
	\begin{frontmatter}
		
		
		
		\title{Effect of rewiring for a sandpile model on a directed network}
		
		
		\author{Alejandro Zamorano} 
		
		\affiliation{organization={Departamento de Física, Facultad de Ciencias, Universidad de Chile},
			city={Santiago},
			country={Chile}}
		
		\author{V\'ictor Muñoz} 
		
		\affiliation{organization={Departamento de Física, Facultad de Ciencias, Universidad de Chile},
			city={Santiago},
			country={Chile}}
		
		\begin{abstract}
			Several studies have considered sandpile dynamics not over regular
			grids, but over networks. In this case, avalanches redistribute grains
			not between neighboring sites in a geometrical sense, but between
			connected sites, in a topological sense. However, depending on how
			nodes are connected, grains may never leave the system, preventing
			energy release. In this work, we study the simplest case, the BTW
			model in one and two dimensions, rewiring the nodes so that at every
			rewiring step, the energy release is always possible, and study
			avalanche statistics as a function of rewiring. In the 1D case, a
			transition is observed in the Gini coefficient of the load
			distribution per node at about 85\% the number of possible rewirings,
			a transition which is not evident with other measures, such as the
			size distribution of avalanches or the mean distance between nodes in
			the network. In the 2D case, energy release follows a power law even
			when the grid is fully rewired, while the Gini coefficient, unlike the
			1D case, decreases at a steady rate, with a smoother transition. The
			effect of network size $N$ is studied, finding that there is a
			transition for the Gini coefficient at the thermodynamic limit $N\to
			\infty$ for both the 1D and 2D cases, transition which is also
			observed in the betweenness centrality, but not in other topological
			measures. Finally, the dependence of the results with the load per
			rewiring iteration, and the avalanche threshold is studied. 
		\end{abstract}
		
		
		
		\begin{keyword}
			Complex Networks
			\sep Self-Organized Criticality
			\sep Nonlinear Dynamics
			
			
			
		\end{keyword}
		
	\end{frontmatter}
	
		
		
		\section{Introduction}
		\label{introduction}
		Self-organized criticality (SOC) provides a fundamental framework for understanding the emergence of complexity in dynamical systems. Sandpile models~\cite{Bak} are the paradigmatic representation of SOC behavior. Traditionally, these models consider a regular grid of cells where an external driver slowly accumulates energy. When the load on a given cell reaches a specific threshold, it is redistributed to neighboring cells, triggering a cascade of relaxations (avalanches) until all loads are below the threshold.
		
		The ubiquity of SOC is reflected in its applications across vastly different natural and social phenomena. Beyond the well-known energy release processes in magnetized space plasmas such as the Earth's magnetosphere~\cite{Chapman} and solar flares~\cite{Strugarek,Charbonneau,lu1993solar,Aschwanden_b} SOC signatures are found in numerous macroscopic systems. For instance, in geomorphology, the flow transported through river networks exhibits directed SOC dynamics via sediment cascades~\cite{fonstad2003self,rinaldo2014evolution}. Similarly, the accumulation and stochastic release of magma in volcanic eruptions follow a directed flow of self-organized criticality~\cite{cannavo2016possible}. Other examples include fractal electrical discharges, where intracloud dynamics release energy following power-law distributions~\cite{iudin2003fractal}, and economic and financial systems, where fluctuations in buying and selling occasionally trigger critical events such as market crashes~\cite{bouchaud2024self}, .
		
		While traditional sandpiles operate on Euclidean lattices, the
		real-world systems such as those mentioned above, are often characterized by heterogeneous, complex structures. Consequently, the generalized case of SOC dynamics over complex networks has gained significant attention~\cite{Lee_b,Ouyang}. Numerical and theoretical studies have extensively explored sandpile models on specific static topologies, such as random~\cite{Bonabeau,Lise}, scale-free~\cite{Goh,Lee_b}, and Apollonian networks~\cite{Vieira}, as well as broader theoretical developments~\cite{Noel}. It is widely accepted that network topology fundamentally modifies avalanche statistics and energy dissipation mechanisms.
		
		A particularly challenging scenario arises when the underlying
		topology is not static but has its own dynamics. Prime examples of such phenomena are found in the aforementioned magnetized plasmas, where avalanches are closely tied to magnetic reconnection processes that continuously modify the topology of the magnetic field~\cite{Uritsky_b,Cargill,Uritsky_a,Abramenko_b}. Recently, we investigated the Lu-Hamilton cellular automaton model for solar flares~\cite{zamorano2025lu}, demonstrating how progressive network rewiring introduces a macroscopic transition in avalanche statistics, leading to a nonlocal development of dissipation events.
		
		In addition to dynamical modeling, complex network analyses for
		magnetized plasmas have proven invaluable for studying the emergent
		topology resulting from observational data, such as solar
		flares~\cite{Gheibi,Najafi}, or sunspots and active
		regions~\cite{Munoz_az,castillo2025observation}. It has even been
		proposed that these observational network methods can be applied to
		forecast the solar cycle~\cite{flandez2025prediction} and to reveal
		the underlying 22-year Hale cycle~\cite{flandez202522}. Whereas
		these data-driven approaches involve building  networks from observed
		events, there remains the issue of how transport properties evolve as
		the topology of the underlying network changes. 
		
		Motivated by this, the present work aims to study the structural
		transition of a simple directed sandpile model as it evolves from a
		regular grid into a complex network. A challenge when selecting an
		arbitrary underlying network to model a SOC system, where energy is
		loaded and dissipated, is the potential emergence of isolated nodes
		(which indefinitely accumulate grains) or closed topological loops
		(which trap energy and prevent global dissipation). To overcome this,
		we apply a controlled rewiring process to the sandpile model, one link
		at a time, ensuring that energy release pathways are always
		mathematically guaranteed. We systematically investigate the resulting
		features of the stationary state of the sandpile, studying the
		distribution of load 
		across the network by means of the Gini coefficient, and charactering
		the network by means of structural metrics such as degree and
		betweenness centrality. Ultimately,
		we apply size scaling analysis to demonstrate how these
		topological alterations trigger a phase transition in the macroscopic
		dissipation of the system. 
		
		The paper is organized as follows: in Sec.~\ref{model}, the directed
		sandpile model and the network construction process are described. The
		numerical and topological results are presented in Sec.~\ref{results},
		followed by an analysis of size effects in Sec.~\ref{finite_size}. Finally, our findings are summarized and discussed in Sec.~\ref{summary}. In addition, robustness to threshold modifications and driving load is presented in the Supplementary Material.
		\section{Model}
		\label{model}
		
		We consider the sandpile model in one and two dimensions, as explained
		below.
		
		\subsection{1D Model}
		\label{1d}
		
		We start from the basic proposal by Bak {\it et al.\/}~\cite{Bak},
		where grains are loaded on a grid, and avalanches occur when the
		height difference between neighboring sites reaches a certain
		threshold $U$. We consider a simple one-dimensional grid of
		$N$ sites, 
		at each timestep $Q=1$ grains are loaded on the grid, on a site which
		is selected randomly.
		
		Since the problem is analogous to a system subject to a constant
		gravitational potential, one can define the energy of the $i$-th site,
		with $n_i$ 
		grains, as
		\begin{equation}
			\label{Ei}
			E_i = \frac{n_i(n_i-1)}2 \ ,
		\end{equation}
		the condition to trigger an avalanche as
		\begin{equation}
			\label{threshold}
			E_i-E_{i+1} \geq U \ ,
		\end{equation}
		the total energy of the system as
		\begin{equation}
			\label{E}
			E=\sum_{i=1}^N E_i \ ,
		\end{equation}
		and the dissipated energy in an avalanche as the difference between
		the energy in the sandpile after $E_\text{after}$ and
		before the avalanche $E_\text{before}$,
		\begin{equation*}
			\Delta E = E_\text{after}-E_\text{before} \ . 
		\end{equation*}
		
\begin{figure}[H]
			\centering
			\includegraphics[width=\textwidth]{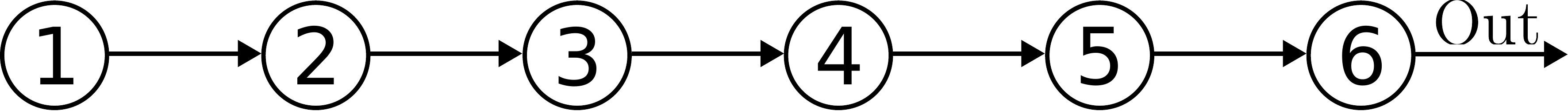}
			\caption{Network representation of the BTW model, where each
				node $i$ delivers energy to node $i+1$, and node $n=6$
				releases energy from the system.} 
			\label{linear}
		\end{figure}

		To study the sandpile dynamics over a network, we first
		represent the system described above with a directed complex network, as seen in
		Fig.~\ref{linear}. Here, each node represents a site in the grid, and
		directed edges represent the direction of the avalanche once the
		threshold condition is met. Thus, 
		the node $i$ is connected to the node $i+1$, so that, if the threshold
		condition is satisfied at node $i$, then a grain is moved from node
		$i$ to node $i+1$. The last node, which has no outgoing
		edges, is the node where energy is finally released from the system
		during the avalanche. A grain loaded into the system, once the
		avalanche occurs, eventually leaves the system at this node. 
		
		This network can be represented by an adjacency matrix full of
		zeros, except for ones above the diagonal. For instance, if the
		network has $N=6$ nodes,  the adjacency matrix would be
		$$M_0=\begin{pmatrix}
			0&1&0&0&0&0\\
			0&0&1&0&0&0\\
			0&0&0&1&0&0\\
			0&0&0&0&1&0\\
			0&0&0&0&0&1\\
			0&0&0&0&0&0
		\end{pmatrix} \ .
		$$
		
		Given this configuration, at each iteration a random node $i$ is
		selected, then one grain is added to it, and the threshold
		condition~\eqref{threshold} 
		is tested. If it is satisfied, one grain is moved in the direction of
		the outgoing edge at $i$, that is, from node $i$ to node $i+1$, and
		the avalanche proceeds until Eq.~\eqref{threshold} is false at every
		site. 
		
		Running a sandpile on this configuration leads of course the same
		results as the BTW model, as shown in Sec.~\ref{results}. However, we
		intend to consider the case where the topology of the network changes,
		so that grain displacement does not necessarily occur on consecutive
		nodes, or equivalently, where adjacent nodes are defined by the
		network connections. To investigate how topological transformations alter the system's configuration and avalanche statistics, we systematically modify the network by rewiring one link at a time until all connections have been updated. This process is executed sequentially: we first rewire the outgoing connections of the first node, then proceed to the second, and so forth. For each rewiring operation, the new target node is chosen uniformly at random, subject to a strict constraint: its index must be greater than that of the original target node.
		
		This topological restriction is critical, as arbitrary rewiring would
		destroy the fundamental properties required for the SOC dynamics. For
		instance, in the linear chain depicted in Fig.~\ref{linear}, if the
		directed edge from node 5 to node 6 were replaced by an edge pointing
		backwards to node 4, then node 6 would be isolated, meaning that grains dropped elsewhere in the network could never reach it. Furthermore, such backward connections introduce closed topological cycles (loops). Consequently, an avalanche reaching node 4 could circulate indefinitely within this loop, trapping the energy and preventing the system from properly dissipating it through the boundaries.
		
		In our treatment, we select the nodes starting from the first node in
		the chain, down to the second to last node in the change, and
		rewire it to a randomly selected node, in such a way that the
		energy can always be released from the grid, and there are
		no accumulation points. This is done by following two rules: (a)
		an edge is modified only once, and (b)
		if the old edge goes from node $i$ to $i+1$, the new edge goes
		from node $i$ to a node $j>i+1$. This is enough to ensure that the
		grid is never broken.

\begin{figure}[H]
			\centering
			\includegraphics[width=\textwidth]{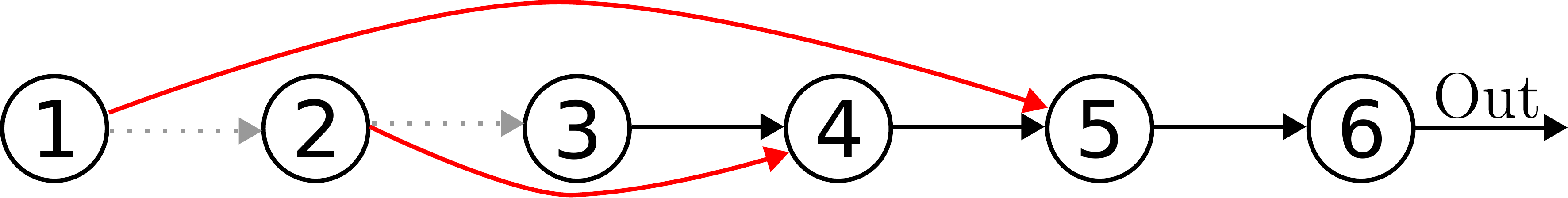}
			\caption{Rewired network configuration and its corresponding
				$M_2$ matrix, after two rewirings. }
			\label{reconnected}
		\end{figure}

		As an example of a rewired grid, see
		Fig.~\ref{reconnected}, where two rewirings have been performed,
		with the resulting adjacency matrix 
		\begin{equation}
			M_2=\begin{pmatrix}
				0&0&0&0&1&0\\
				0&0&0&1&0&0\\
				0&0&0&1&0&0\\
				0&0&0&0&1&0\\
				0&0&0&0&0&1\\
				0&0&0&0&0&0
			\end{pmatrix} \; .
		\end{equation}
		
		Regardless of where grains are added to the sandpile, energy is always
		released at node 6. 
		
		In general, valid rewirings, which do not separate the system into
		isolated subnetworks, and always have a single node where energy
		leaves the system, can
		be easily represented in terms of the adjacency 
		matrix as follows:
		\begin{equation}
			\label{rule}
			\begin{aligned}
				M_{i,i+1} &= 0\; ,\\
				M_{i,\text{rnd}[i+1,N]}&=1 \; ,
			\end{aligned}
		\end{equation}
		where node $i$ has been selected for rewiring, and
		$\text{rnd}[i+1,N]$ represents a random integer $j$ such that $i+1
		< j \leq N$. The initial linear configuration of $N$ nodes, can be
		rewired at most $N-2$ times. \\
		We can see the psuedocode in Alg.~\ref{alg:rewiring}, 
		
		\begin{algorithm}[H]
			\caption{Directed Rewiring Algorithm (Preserving Direct Acyclic Graph (DAG) Structure)}
			\label{alg:rewiring}
			\begin{algorithmic}[1]
				\Require Initial Network $G=(V, E)$ (Directed Grid), Target Rewiring Count $n_{target}$
				\Ensure Rewired Network $G'$ (Directed Acyclic Graph, constant total degree)
				
				\State $count \gets 0$
				\While{$count < n_{target}$}
				\State Select a source node $i \in V$ uniformly at random
				\If{$i$ is a sink node (boundary)}
				\State \textbf{continue}
				\EndIf
				
				\State Identify the current neighbor $k$ such that $(i, k) \in E$
				\State Select a candidate target node $j>i \in V$ uniformly at random
				
				\Comment{Constraint 1: Avoid self-loops and existing connections}
				\If{$i \neq j$ \textbf{and} $j \neq k$}
				
				\Comment{Constraint 2: Ensure Acyclicity (DAG)}
				\If{There is no directed path from $j$ to $i$}
				\State $E \gets E \setminus \{(i, k)\}$ \Comment{Remove old link}
				\State $E \gets E \cup \{(i, j)\}$    \Comment{Add new link}
				\State $count \gets count + 1$
				\EndIf
				\EndIf
				\EndWhile
				\State \Return $G(V, E)$
			\end{algorithmic}
		\end{algorithm}
		The last node does not have an outgoing edge, and every grain reaching it is discarded
		unconditionally, whereas the second to last
		node has only one connection (to the last node), so both nodes cannot
		be rewired.
		
		Notice that this algorithm generates all possible directed acyclic
		graphs with $N$ nodes, such that $N-1$ nodes have out-degree equal to
		1. (The exception is the last node.)  
		More general acyclic graphs could be built by adding new
		connections instead of only rewiring, or by more general
		algorithms~\cite{Oden}, but here we focus on incremental departures
		from the original BTW model. Also, by construction, at every step in
		the algorithm the resulting network has a tree structure, two
		arbitrary 
		nodes being connected by a unique path. This tree-like structure
		arises naturally, as all paths must reach the last node where energy
		is released, and all connections must be directed from a certain $n$-th
		node to an $m$-node, with $m>n$.
		
		The resulting network structure is also interesting, as various
		systems which exhibit SOC behavior, also involve direct transport on
		a fractal-like structure  with a shape similar to directed acyclic
		graph, as energy flows to a sink. Such is the case, for instance, of
		river basins, magma flows, and electrical discharges
		\cite{pelletier1999self, tarboton1988fractal,shaw1991fractal,
			iudin2003fractal}. The concept of directed acyclic graphs has also
		gained traction in economics and finance, where they are used to model
		causal structures in price discovery and risk propagation
		\cite{zema2022directed,haigh2004causality,d2021evolving}. 
		
		In principle, the rewiring algorithm described above could be used for
		any type of network, 
		preserving the 
		number of connections and nodes. However, in our case, outgoing
		nodes are not changed from the initial configuration, then a random
		choice of nodes to rewire can break
		the network, affecting energy release. Further improvements could be
		made by accepting rewirings only if they do not break the network, or
		adding new outgoing modes, but we will consider the simple model
		proposed to study it systematically.

		Now, given a certain rewired configuration, we iterate the
		sandpile model, adding grains at random nodes at each iteration, and
		study the features of the resulting stationary state as the underlying
		grid departs from the simple linear chain, as a function of the number
		of rewirings.

		\subsection{2D Model}
		\label{2d}
		
		We start from the unperturbed adjacency matrix is 
		\begin{equation}
			\label{M0_2d}
			M_0=\begin{pmatrix}
				0&1&0&1&0&0&0&0&0\\
				0&0&1&0&1&0&0&0&0\\
				0&0&0&0&0&1&0&0&0\\
				0&0&0&0&1&0&1&0&0\\
				0&0&0&0&0&1&0&1&0\\
				0&0&0&0&0&0&0&0&1\\
				0&0&0&0&0&0&0&1&0\\
				0&0&0&0&0&0&0&0&1\\
				0&0&0&0&0&0&0&0&0
			\end{pmatrix} \ ,
		\end{equation}
		which is equivalent to the 2D grid of the BTW model, as shown in
		Fig.~\ref{grid}: 
		\begin{figure}[H]
			\centering
			\includegraphics[width=0.6\textwidth]{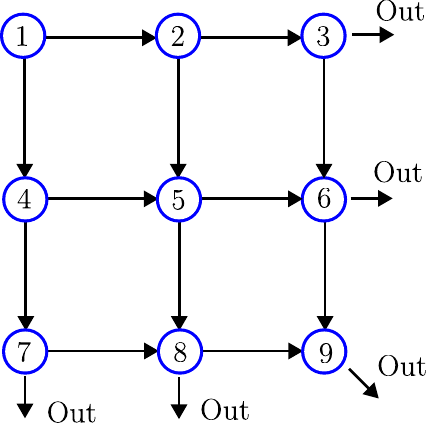}
			\caption{Graphical representation of the initial adjacency
				matrix \eqref{M0_2d}.}
			\label{grid}
		\end{figure}
		
		In this case, energy leaves the system at the edges. Following the
		strategy for the 1D model, we now rewire nodes in such a way that
		energy always flows downhill, and that 
		energy can be released from the system at the same nodes as in the
		regular 2D grid. This means that
		connections leaving the network are not modified. 
		
		Thus, the same two rewiring rules as in the 1D case, Eq
		(\ref{rule}), can be used.

		For instance, after two rewirings the system can be given by 
		an adjacency matrix such as the following:
		\begin{equation}
			\label{M2_2d}
			M_2=\begin{pmatrix}
				0&0&0&0&1&1&0&0&0\\
				0&0&1&0&1&0&0&0&0\\
				0&0&0&0&0&1&0&0&0\\
				0&0&0&0&1&0&1&0&0\\
				0&0&0&0&0&1&0&1&0\\
				0&0&0&0&0&0&0&0&1\\
				0&0&0&0&0&0&0&1&0\\
				0&0&0&0&0&0&0&0&1\\
				0&0&0&0&0&0&0&0&0
			\end{pmatrix} \ .
		\end{equation}
		This can be represented by Fig.~\ref{grid2}:
		\begin{figure}[H]
			\centering
			\includegraphics[width=0.6\textwidth]{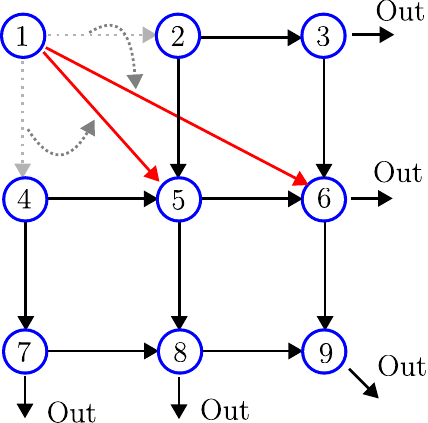}
			\caption{Rewired network configuration according to the
				$M_2$ matrix in Eq.~\eqref{M2_2d}, after two rewirings.}
			\label{grid2}
		\end{figure}
		
		\section{Results}
		\label{results}
		
		\subsection{1D Case}
		\label{results1d}
		
		In the following, we study a system with $N=256$~nodes,
		an energy
		threshold $U=2$ to trigger an avalanche, and where $Q=1$
		grains are loaded on the sandpile at a time. For every run,
		we consider $10^8$ iterations, which is more than enough to reach
		the stationary state. And we follow the system for $10^5$ iterations once the
		transient has ended.  
		
\begin{figure}[H]
			\centering
			\includegraphics[width=\textwidth]{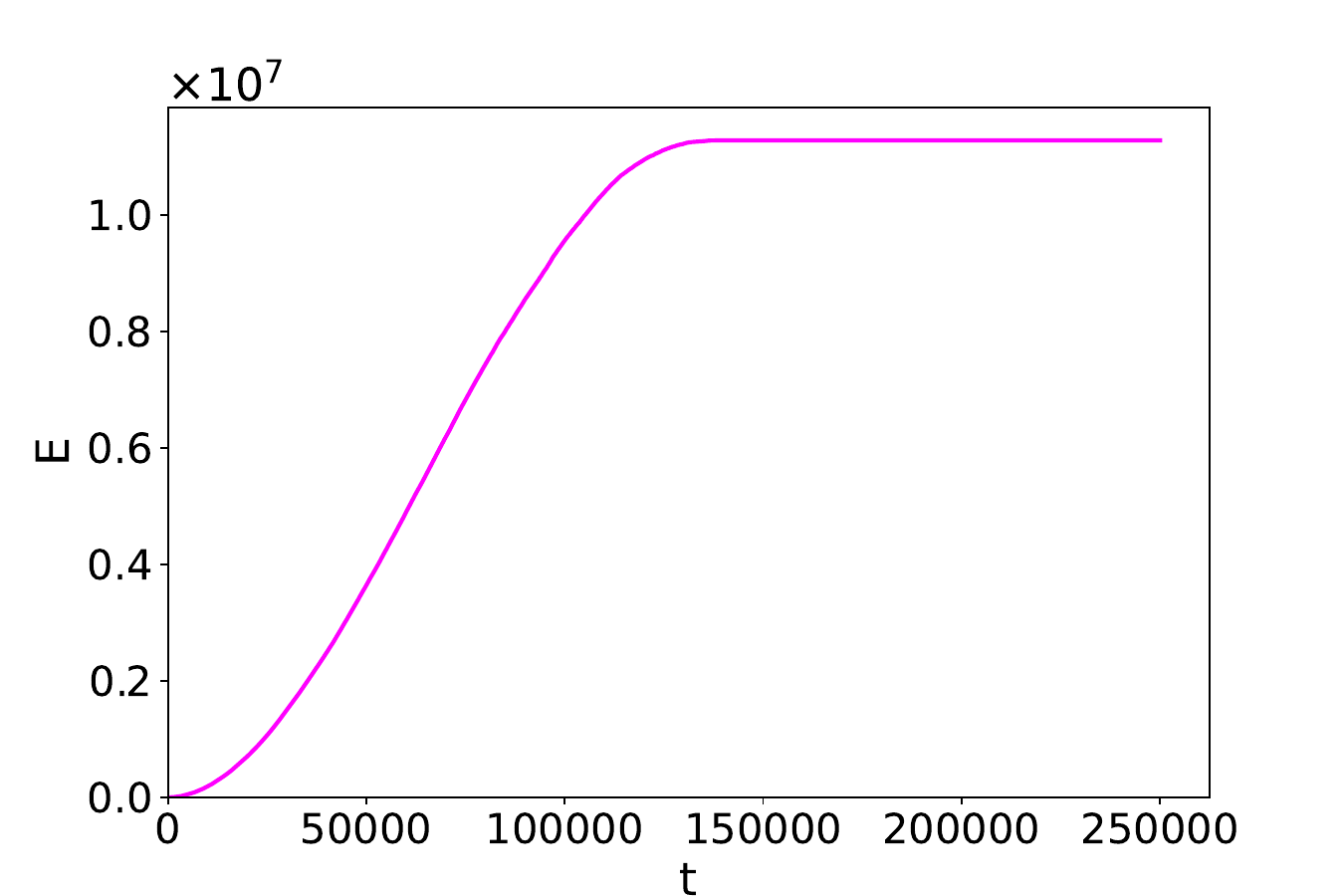}
			\caption{Energy of the system for the non-rewired 1D case
				(linear chain) as a function of iteration time $t$, showing the transition from the transient to
				the stationary state.} 
			\label{energyzoom}
		\end{figure}

		For the initial linear chain, the stationary state is reached after
		about 133\,000 iterations, as shown in Fig.~\ref{energyzoom}, where
		the total energy $E$ in the system is plotted as a function of
		iteration time $t$.
		
\begin{figure}[H]
			\centering
			\includegraphics[width=\textwidth]{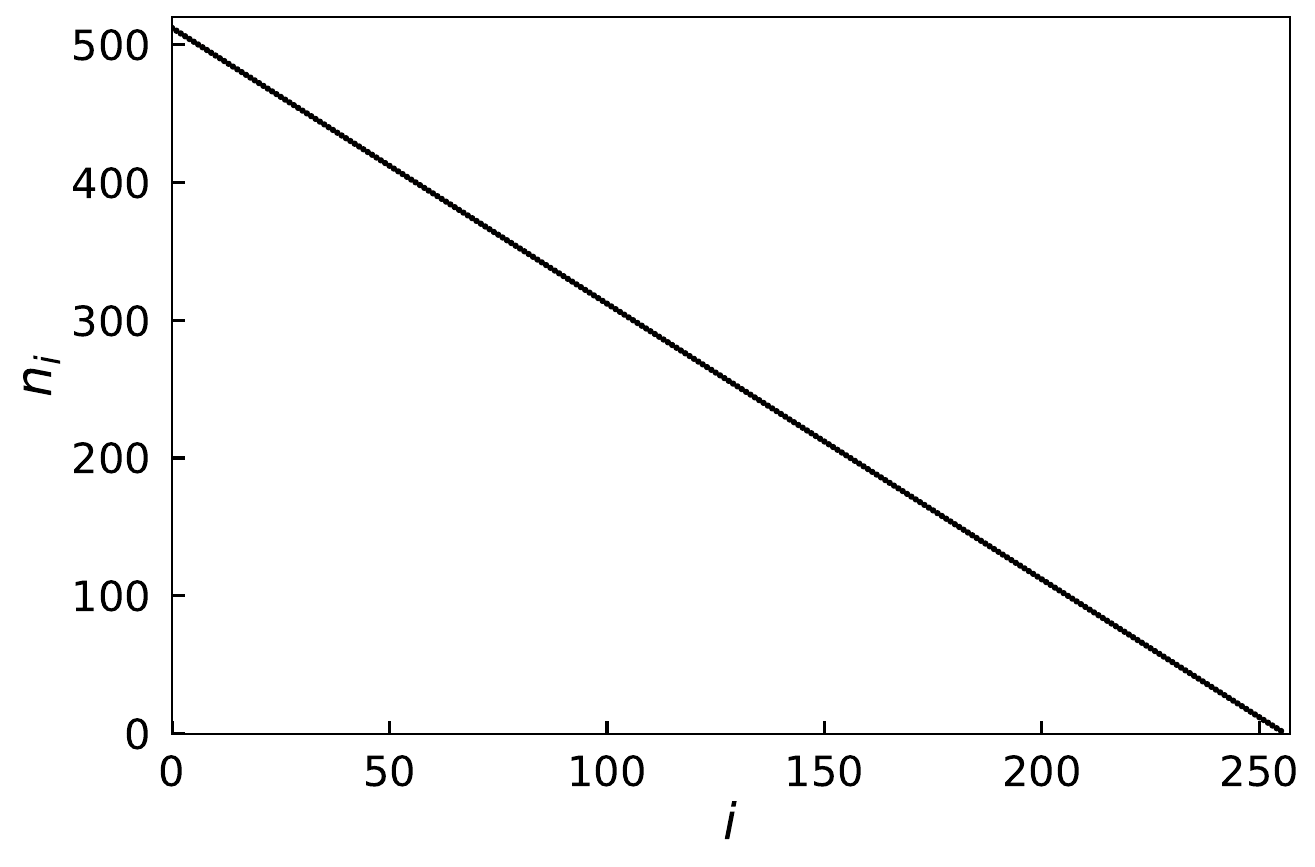}
			\caption{Number of grains at the $i$-th state at the stationary state
				for the linear chain, with no 
				rewiring. }
			\label{linear_state}
		\end{figure}

		The shape of the stationary state is, as expected, a simple configuration
		where the $i$-th node has $U=2$ grains more than the $i+1$-th node
		(Fig.~\ref{linear_state}).
		
\begin{figure}[H]
			\centering
			\includegraphics[width=\textwidth]{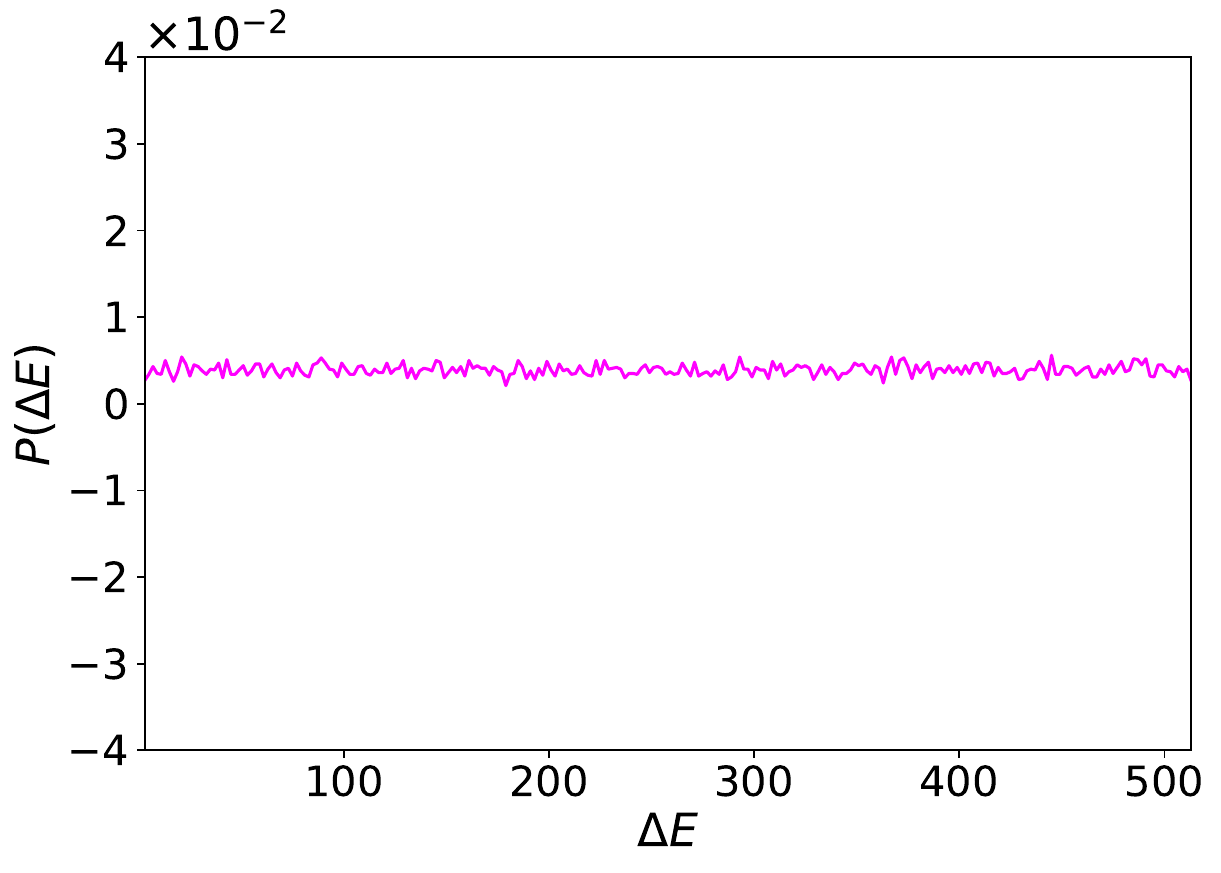}
			\caption{Probability Distribution Function (PDF) of the energy
				released $\Delta E$ during avalanches for the linear chain,  without
				rewirings.} 
			\label{linear_pdf}
		\end{figure}

		The Probability Distribution Function (PDF) of energy release events at the
		stationary state is 
		essentially constant (Fig.~\ref{linear_pdf}) which is consistent with
		the fact that during an 
		avalanche a grain simply moves through the pile until the last site,
		and avalanches of all sizes can occur with equal probability,
		depending only on the site where the grain is loaded, which is itself
		chosen from a uniform distribution. Thus, avalanche size statistics
		are trivial in this case~\cite{Bak}.  
		
		Since the distribution of grains is not uniform across nodes in the
		stationary state, it makes sense to calculate 
		the ``inequality'' of its distribution, through the Gini
		coefficient, which turns out to be $G\simeq 0.33$ for the linear
		chain. This will give us an easy way to characterize the shape of the
		stationary state as rewirings are introduced.
		
\begin{figure}[H]
			\centering
			\includegraphics[width=\textwidth]{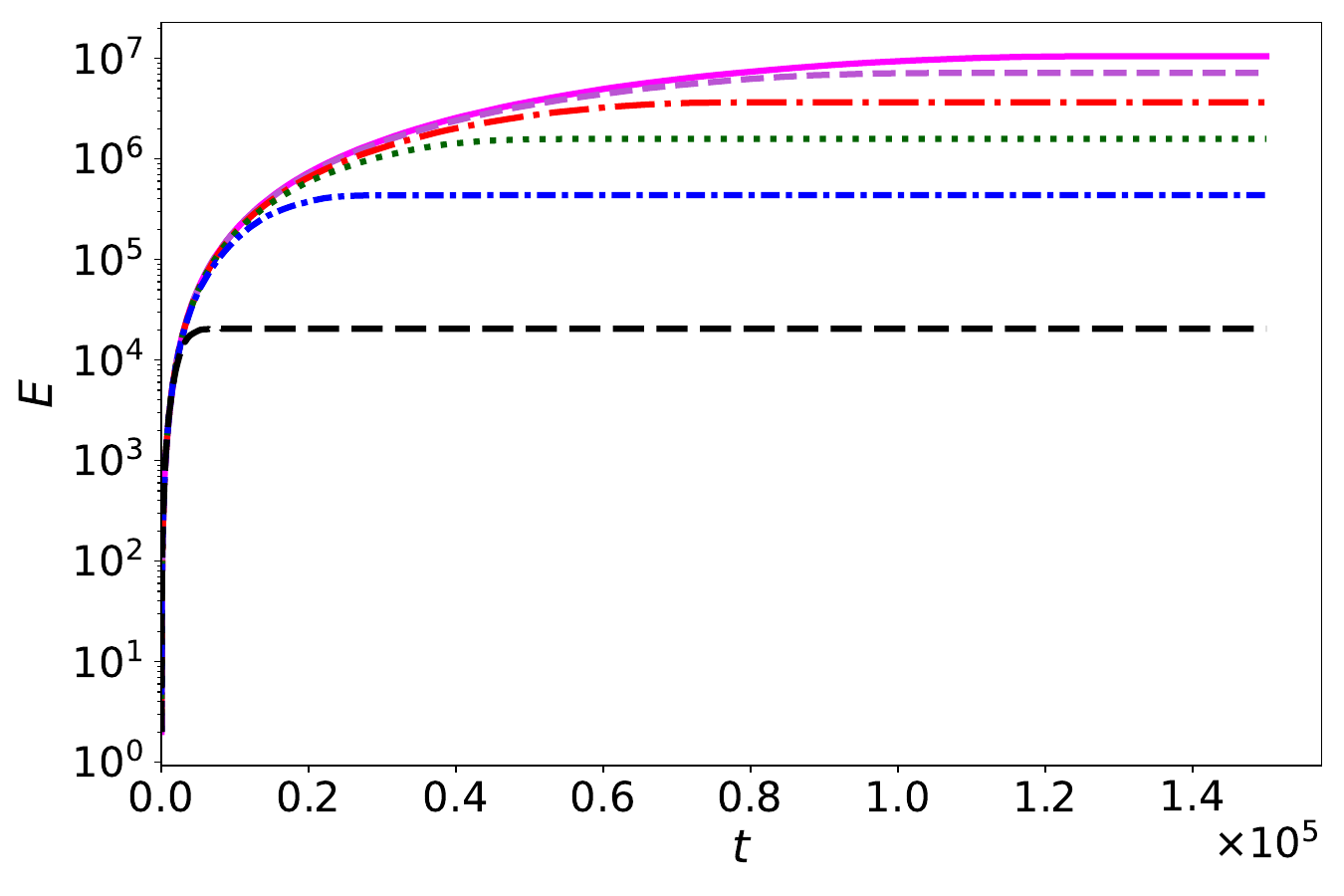}
			\caption{Energy in the sandpile at each iteration for the
				rewired case, with 0 (magenta line), 50 (purple line), 100
				(red line), 150 (green line), 200 (blue line), and 250 (black
				line) rewirings.}
			\label{energy}
		\end{figure}

		We now rewire the sandpile, as described in Sec.~\ref{model}, and
		study the effect on the stationary state. The first observation is
		that, as rewirings are made, the average energy in the system
		decreases, and the number of iterations needed to reach the stationary
		state decreases. This is shown in Fig.~\ref{energy}, where curves of
		the energy in the sandpile after each iteration are shown for six
		different values of the number of rewirings $n_R$.
		
		Before rewirings, all nodes receive avalanches from the previous
		one, and they can
		discharge only to the following node. By
		rewiring, the paths of discharge are increased, which leads to a
		shorter time to reach the stationary state. This also decreases the
		energy of the stationary state: since discharges can follow more than
		one path, the condition \eqref{threshold} must be met by more pairs of
		nodes, and since loads to the network are at random sites, the
		threshold condition can be satisfied earlier than for the linear
		chain, reducing the average load of nodes.
		
\begin{figure}[H]
			\centering
			\subfigure[~0 rewirings.]{\includegraphics[width=0.45\linewidth]{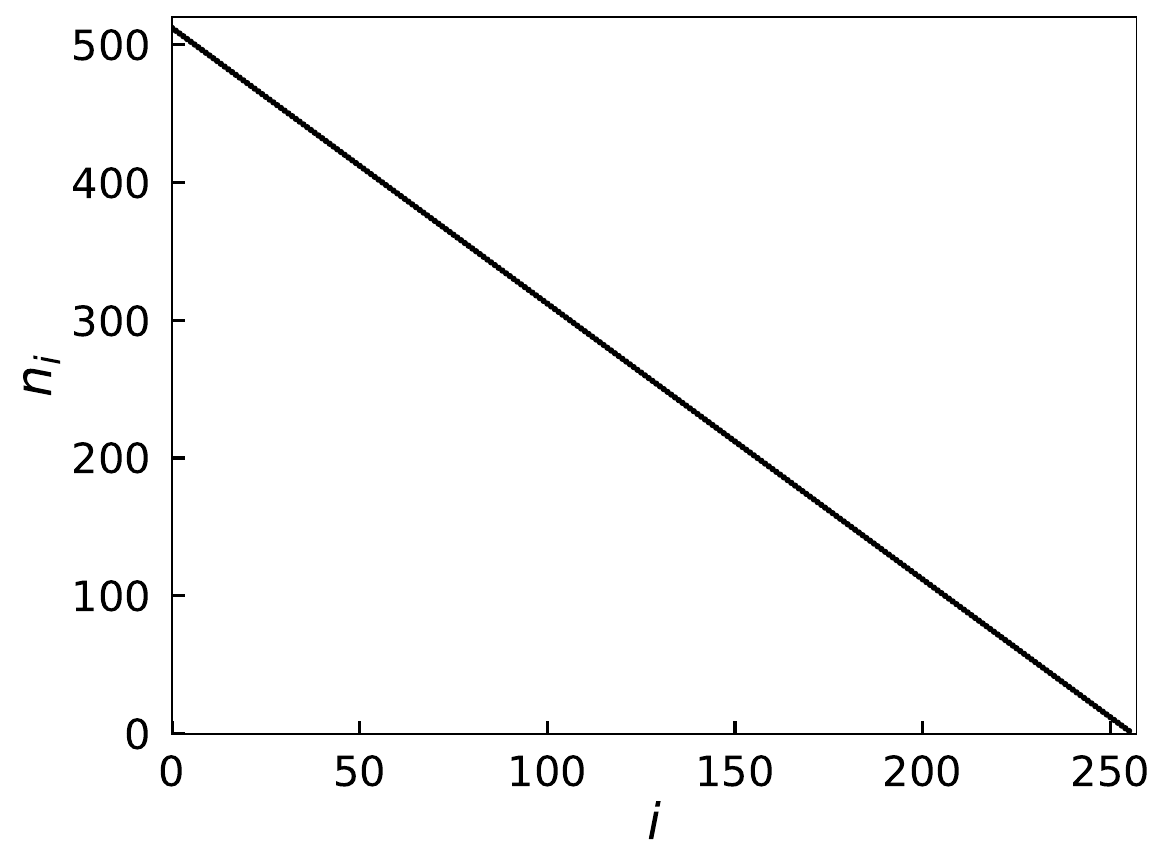}}
			\subfigure[~50 rewirings.]{\includegraphics[width=0.45\linewidth]{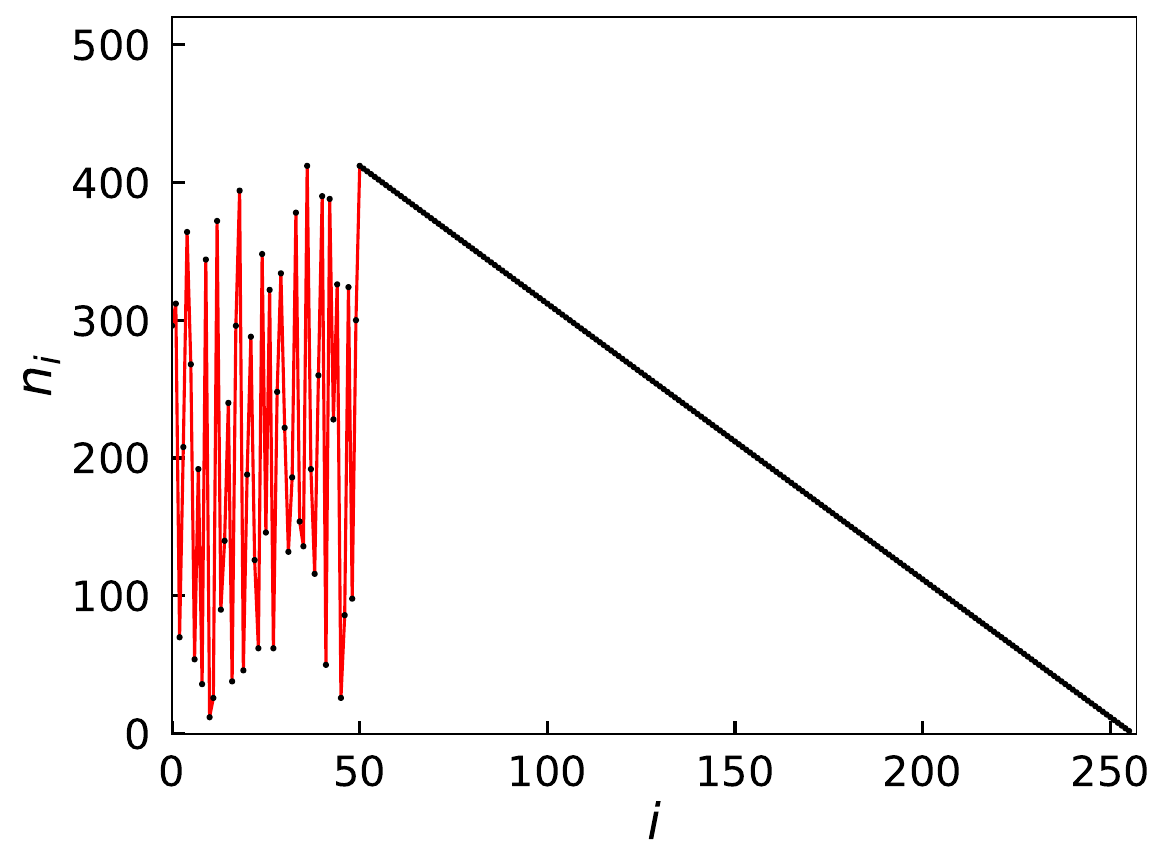}}
			\subfigure[~100 rewirings.]{\includegraphics[width=0.45\linewidth]{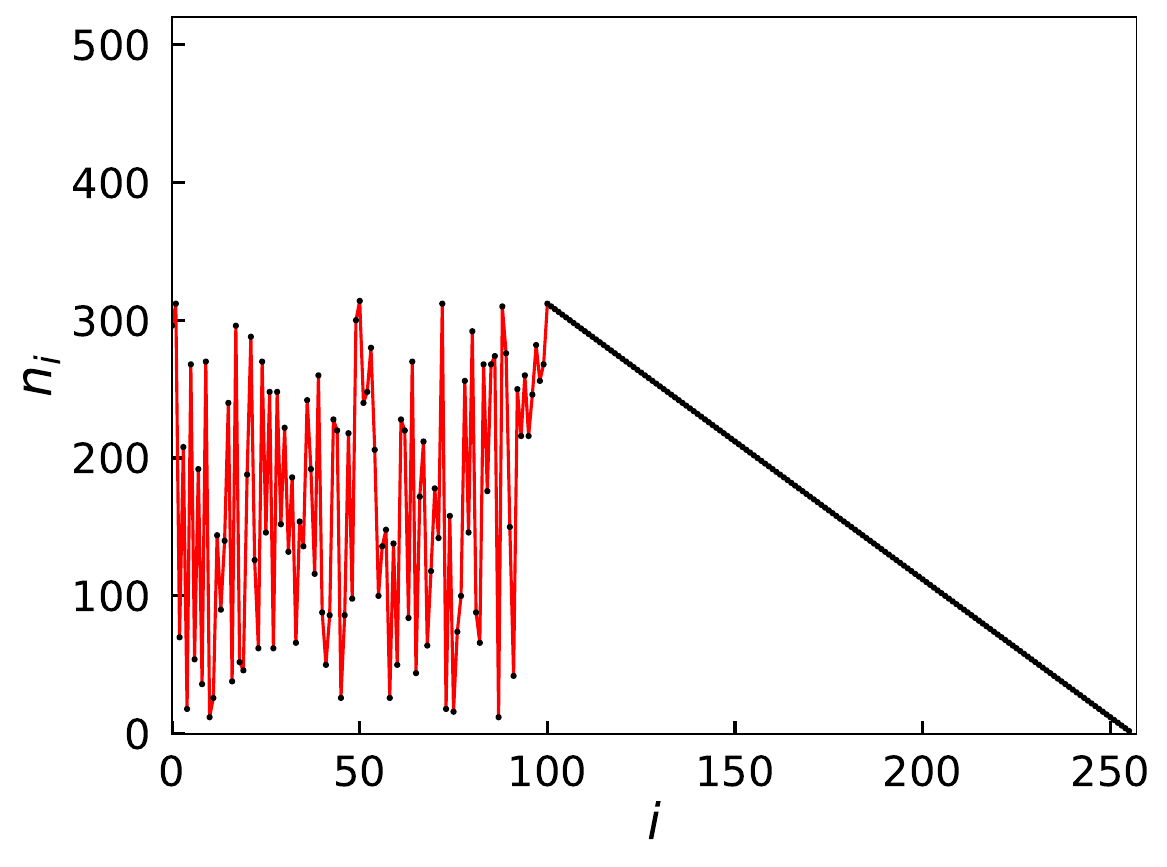}}
			\subfigure[~150 \hspace*{-0.1cm} rewirings.]{\includegraphics[width=0.45\linewidth]{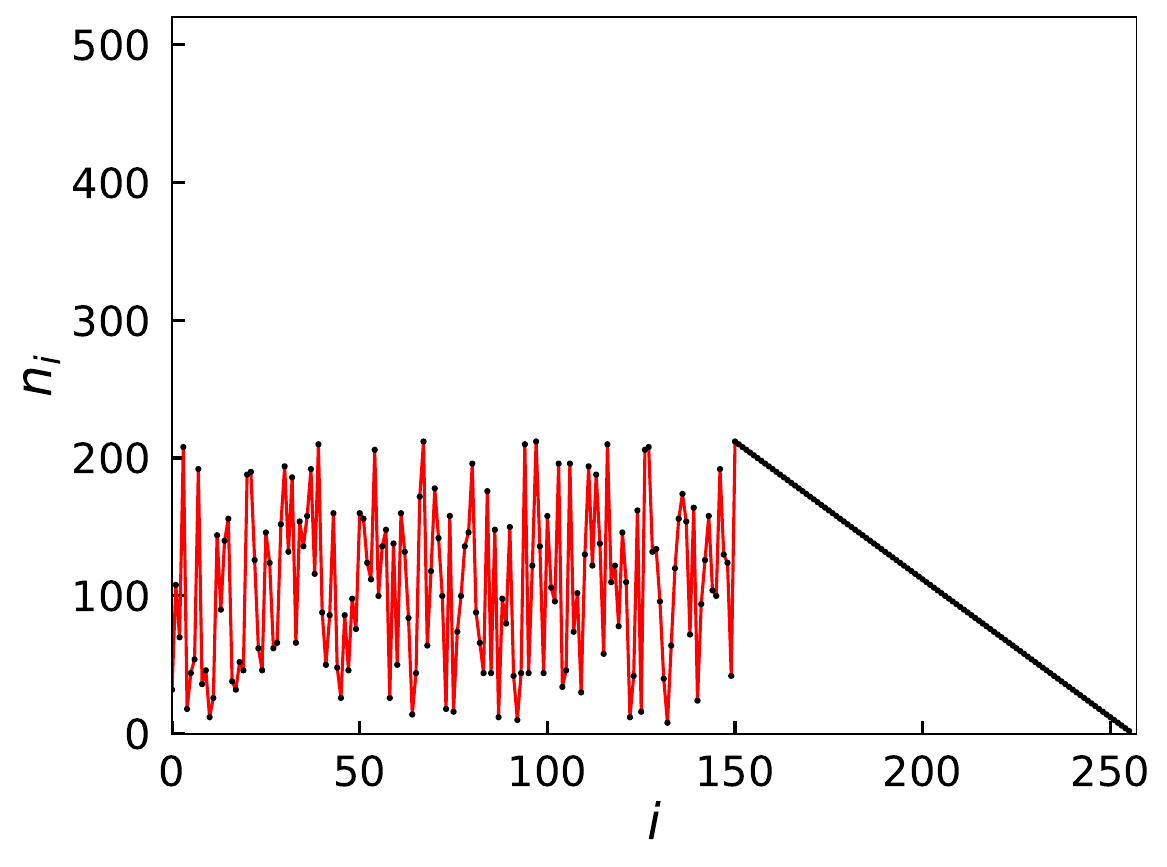}}
			\subfigure[~200 rewirings.]{\includegraphics[width=0.45\linewidth]{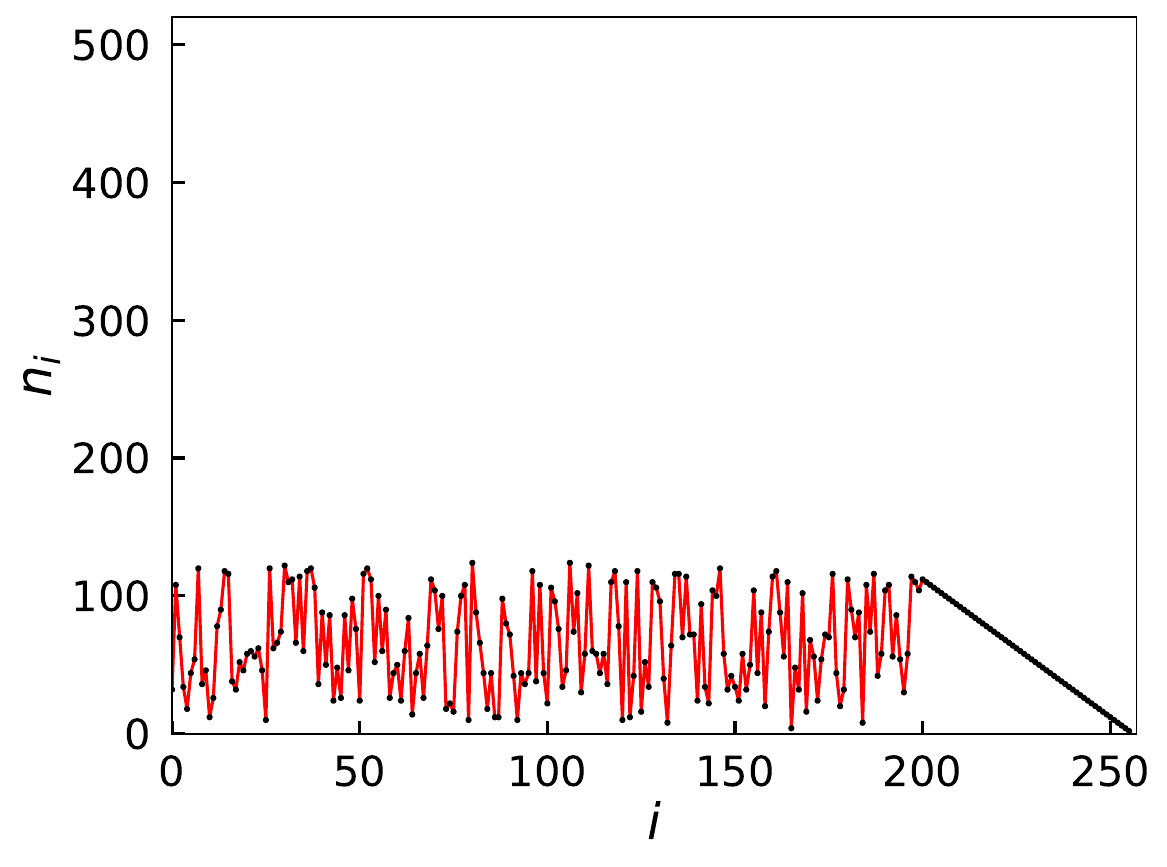}}
			\subfigure[~250 rewirings.]{\includegraphics[width=0.45\linewidth]{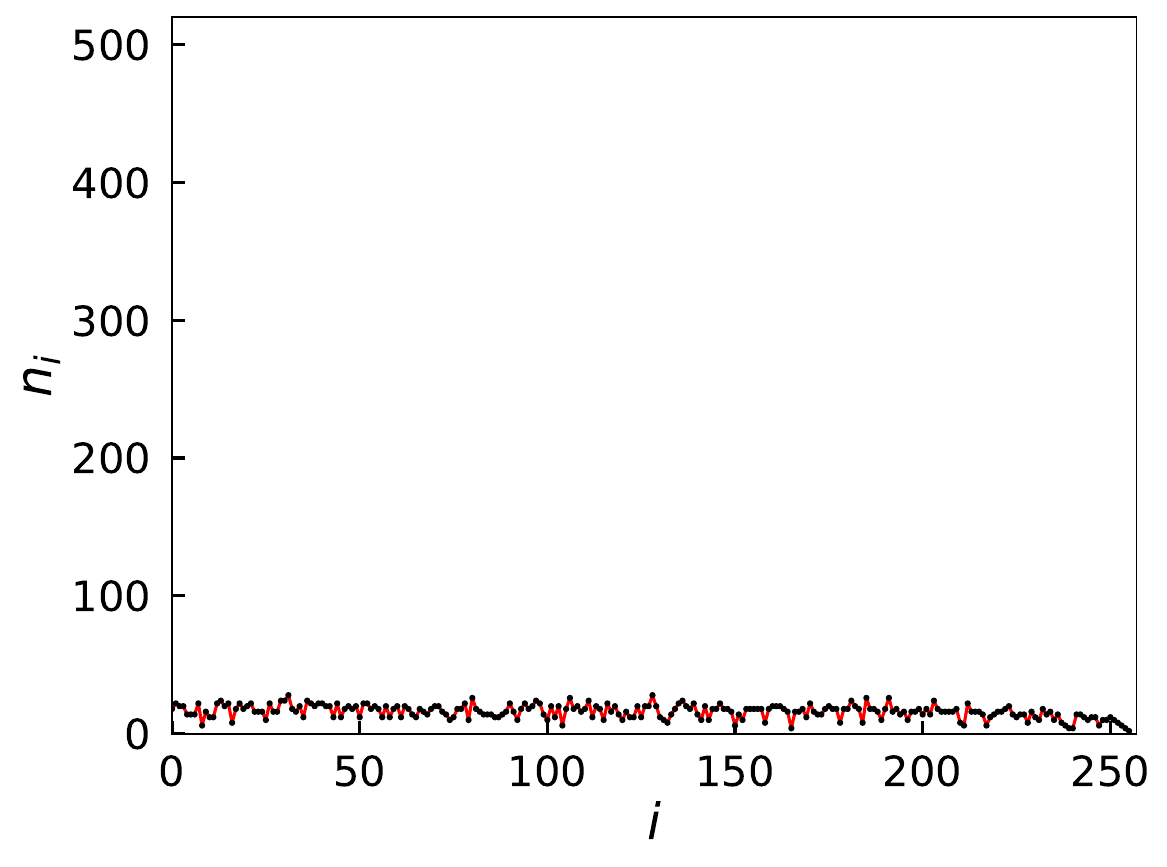}}
			\caption{Stationary states associated to the rewired
				networks, for $U = 2$, and $Q = 1$, and various
				numbers of rewirings:  a 0 rewirings, b) 50, c) 100, d) 150, e) 200 and f) 250.} 
			\label{shape}
		\end{figure}

		This can be clearly seen by observing the shape of the stationary
		states, as shown in Fig.~\ref{shape}, for the same networks
		represented in Fig.~\ref{energy}. 
		
		As expected from the analysis of Fig.~\ref{energy}, rewirings have
		the effect of reducing the average load of all nodes, since discharges
		are more likely to occur. It can also be clearly noticed that this
		occurs first in the first nodes in the chain, given the sequence in
		which nodes are rewired. Thus, rewirings are
		enough to destroy the simple configuration of the stationary state BTW
		model, and which in turn leads to a trivial dynamics, as shown in
		Ref.~\cite{Bak}. In Ref.~\cite{Bak}, a more interesting dynamics is found by
		considering the bidimensional case. Here, we see an alternative way by
		rewiring the linear chain, which is another way to provide more
		discharge paths.
		
		As can be seen in Fig.~\ref{energy}, this change in topology causes
		the total energy of the rewired system to reach the critical state
		sooner; thus, when analyzing the avalanche statistics, the transient
		is always shorter than that of the original BTW system (This applies
		to both 1D and 2D). 
		
\begin{figure}[H]
			\centering
			\includegraphics[width=\textwidth]{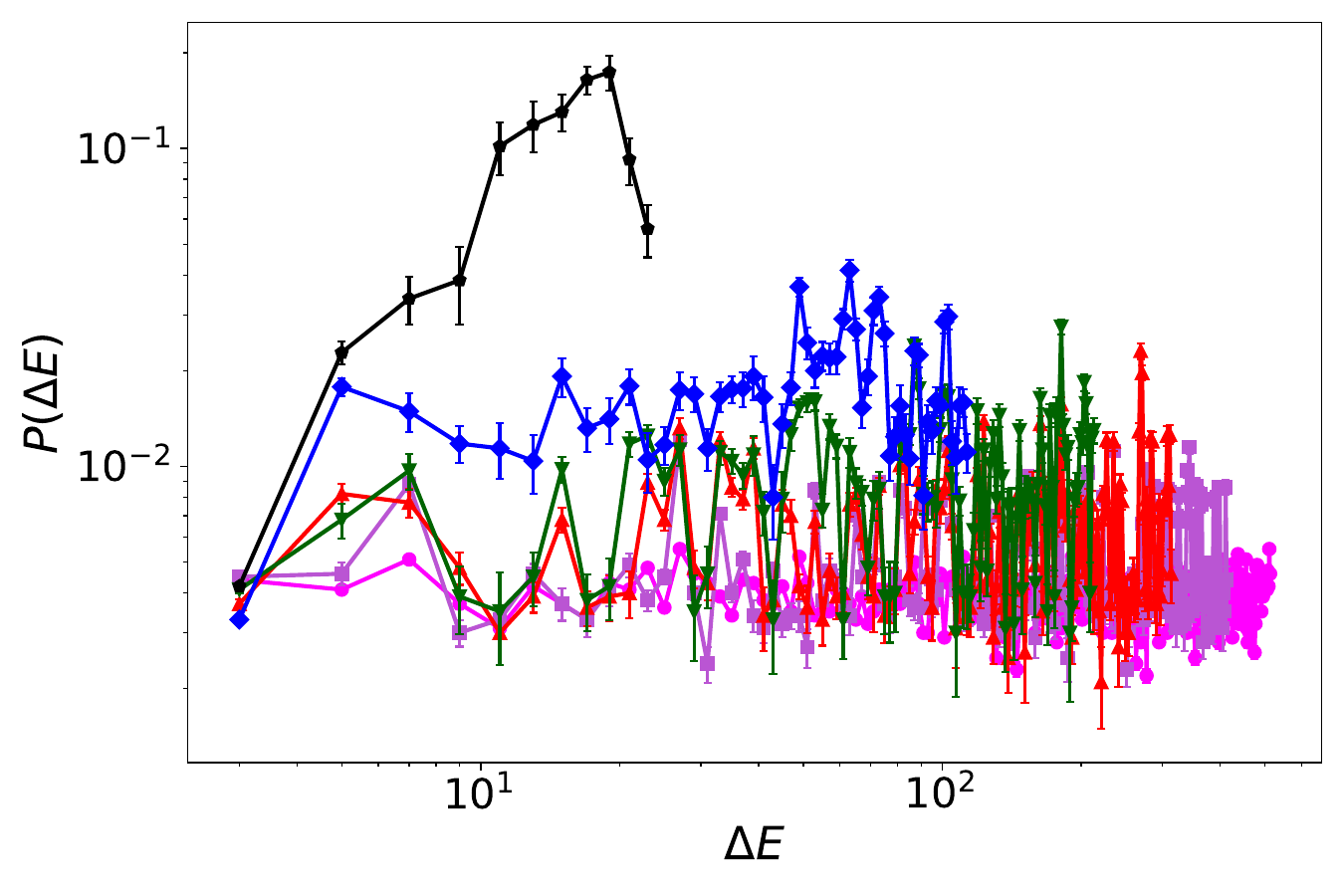}
			\caption{PDF of energy release events for the rewired
				networks, averaged over 30 simulations with the same load
				$Q=1$ and threshold $U=2$, each simulation corresponding to
				a certain sequence of rewirings. Line styles represent
				different numbers of rewirings: 0 (magenta line, with circles), 50 (purple line, with squares), 100
				(red line, with triangles), 150 (green line, with inverted triangles), 200 (blue line, with diamonds), and 250 (black
				line, with pentagons) rewirings.}
			\label{pdf}
		\end{figure}

		The changes in the profile of the stationary states also modify
		avalanche statistics. This is shown in Fig.~\ref{pdf}, where the PDF
		of the energy of avalanche events are plotted, for networks with the
		same number of rewirings as in Figs.~\ref{energy}
		and~\ref{shape}.  Since rewirings are selected randomly, we have
		taken 30 different sequences of rewirings, consistent with the
		rules in Sec.~\ref{model}, and averaged their results. Thus, the curves in
		Fig.~\ref{pdf} are the average over 30 runs of different networks with 0
		rewirings (magenta curve), 50 rewirings (purple curve), and so on.
		
		Consistent with Figs.~\ref{energy} and~\ref{shape}, it can be observed
		that large events are less likely as more rewirings are
		introduced. Also, avalanches occur in a narrower range of energies,
		consistent with the fact that it is no longer possible for a single
		site to accumulate a large energy anyway, as in the simple linear
		chain (Fig.~\ref{shape}(a)).
		
\begin{figure}[H]
			\centering
			\includegraphics[width=\textwidth]{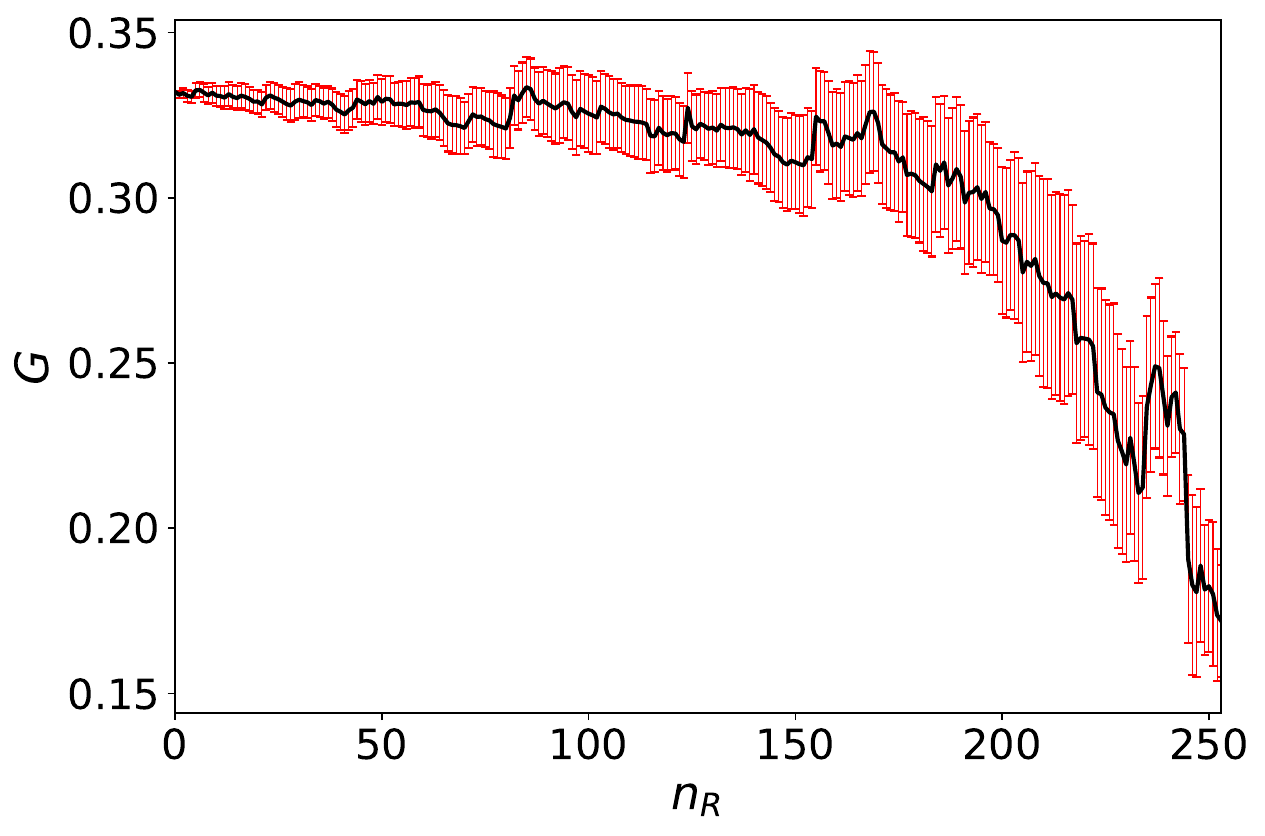}
			\caption{Gini coefficient for the distribution of load in the
				stationary state, averaged over 30 samplings of the random
				sequence of rewiring. Error bars correspond to the
				standard deviation.}
			\label{gini}
		\end{figure}

		The complementary cumulative distribution functions $F(E)$ do not show a
		power-law scaling, consistent with the original BTW model,
		thus this particular feature is not modified by successive rewirings
		(see the Supplementary Material).
		However, one metric that shows a strong dependence on the number of
		rewirings is the Gini coefficient ($G$) of the load across the sandpile.
		This is shown in Fig.~\ref{gini},
		where, again, 30 different random sequences of rewirings were averaged. It
		is clear that, up to 150 rewirings, the inequality of the
		distribution of load in the network is approximately constant, and
		then inequality rapidly decreases. The transition occurs at
		approximately 85\% of the rewiring process. In Sec.~\ref{finite_size}
		this is discussed in more detail, where size effects are considered.
		
\begin{figure}[H]
			\centering
			\includegraphics[width=\textwidth]{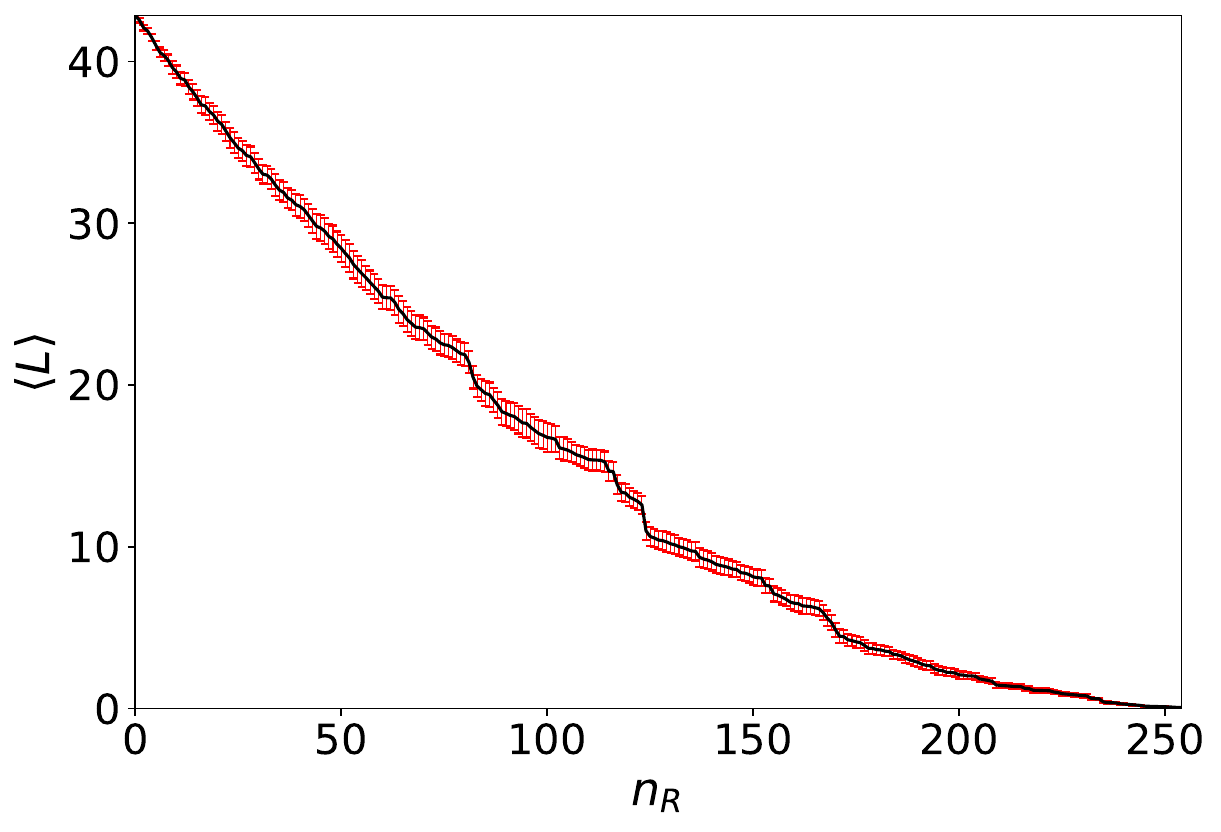}
			\caption{Average distance between nodes as a function of the
				number of rewirings, averaged over 30 sequences of
				rewirings. } 
			\label{distance}
		\end{figure}

		Although the percolation argument makes sense, since more paths are
		available for the discharge, it is interesting to notice that this is
		not necessarily correlated to other measures of connectivity between
		nodes. A simple example is the average shortest path between
		nodes, defined as
		$$
		\langle L\rangle =\sum_{s,t,s\not=t}\frac{d(s,t)}{n(n+1)} \ ,
		$$
		where $d(s,t)$ is the shortest path from node $s$ to node $t$, $n$
		is the number of nodes in the network, and $d(s,t)\equiv0$ if 
		there is no path from $s$ to $t$. As seen in Fig.~\ref{distance},
		the distance between  
		nodes decreases as the number of rewirings increases, which is
		expected, but the
		decrease is much faster than for the Gini coefficient, and no
		particular transition is found. 
		
		We have carried out a similar analysis but changing the threshold to
		all integer values from $U=2$ up to $U=7$, in order to determine how
		results are 
		affected by the specific value of the threshold chosen.
		These results, which consider different load values $Q$ and threshold
		values $U$ to study the robustness of the method, are included
		in the Supplementary Material. 
		In the Supplementary Material shows the
		complementary cumulative distribution 
		functions for various values of the threshold $U$, for load $Q=1$, when the maximum
		number of rewirings has been reached. The main effect is the
		broadening of the distribution, as larger values of the released
		energy in each avalanche are possible, which is expected if the
		load is constant, but a larger threshold is needed to trigger an avalanche. No
		clear exponential or polynomial tails are observed. 
		
		\subsection{2D Case}
		\label{results2d}
		
		For the 2D system, we performed simulations with 256
		nodes (a $16\times 16$ matrix), on which 30 different
		simulations were performed. Each simulation involves 480
		rewirings and $10^5$ iterations. The system starts from
		a steady state, using driver $Q=1$ and threshold $U=3$.

		\begin{figure}[H]
			\centering
			\includegraphics[width=\textwidth]{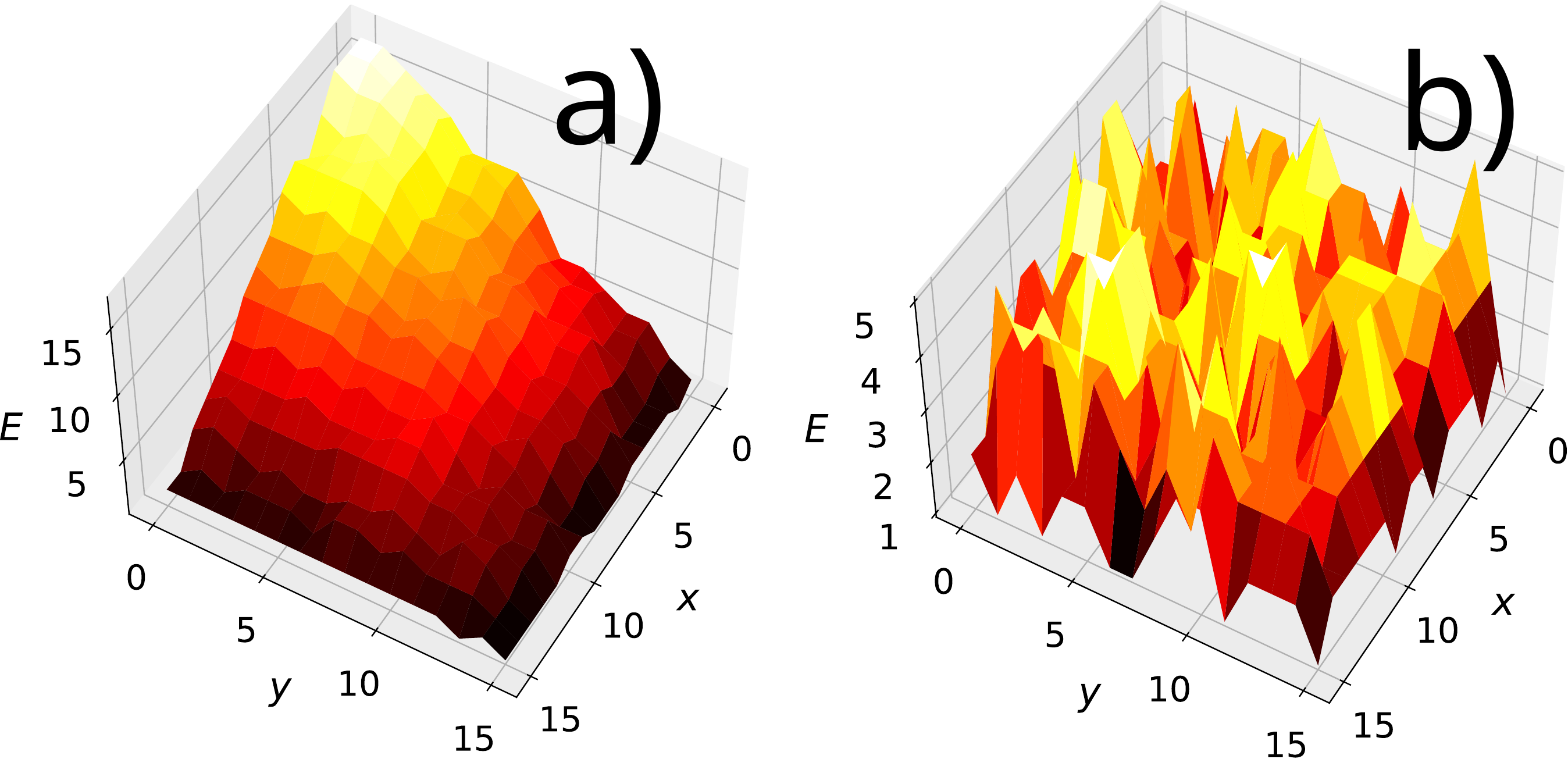}
			\caption{Critical state for the $N=16\times16$ system a) without rewirings,
				b) after 400 rewirings. 
				The color map represents the number of grains per site.} 
			\label{2dmodel}
		\end{figure}

\begin{figure}[H]
			\centering
			\includegraphics[width=\textwidth]{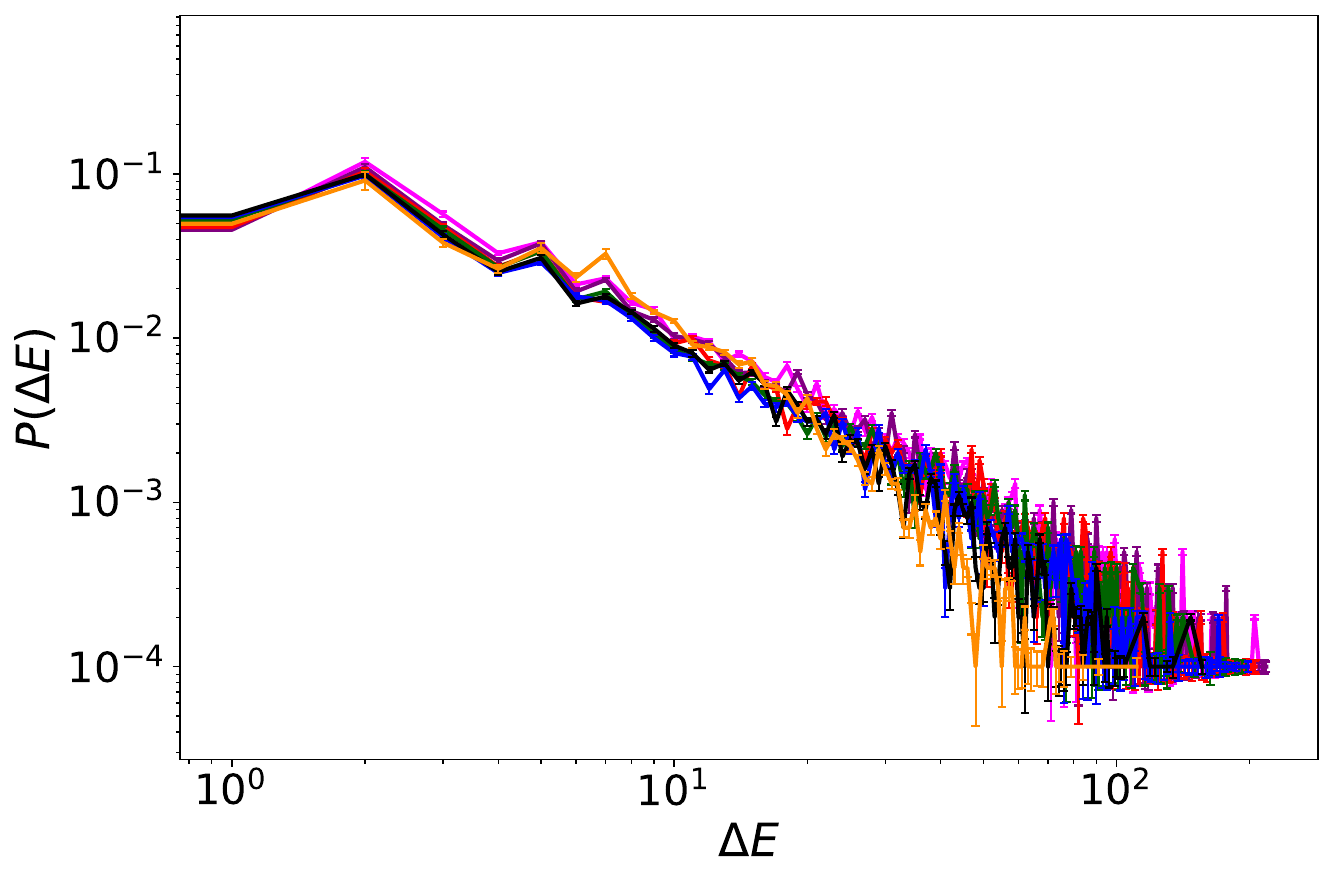}
			\caption{PDF of released energy for
				different rewirings, in logarithmic scale, once the
				stationary critical state is reached. Colors
				correspond to 0 (magenta line), 80 (purple line), 160 (red
				line), 240 (green line), 320 (blue line), 400 (black
				line), 480 (brown line) rewirings. Charge is $Q=1$
				and threshold 
				is $U=3$.}
			\label{pdf2d}
		\end{figure}

\begin{figure}[H]
			\centering
			\includegraphics[width=\textwidth]{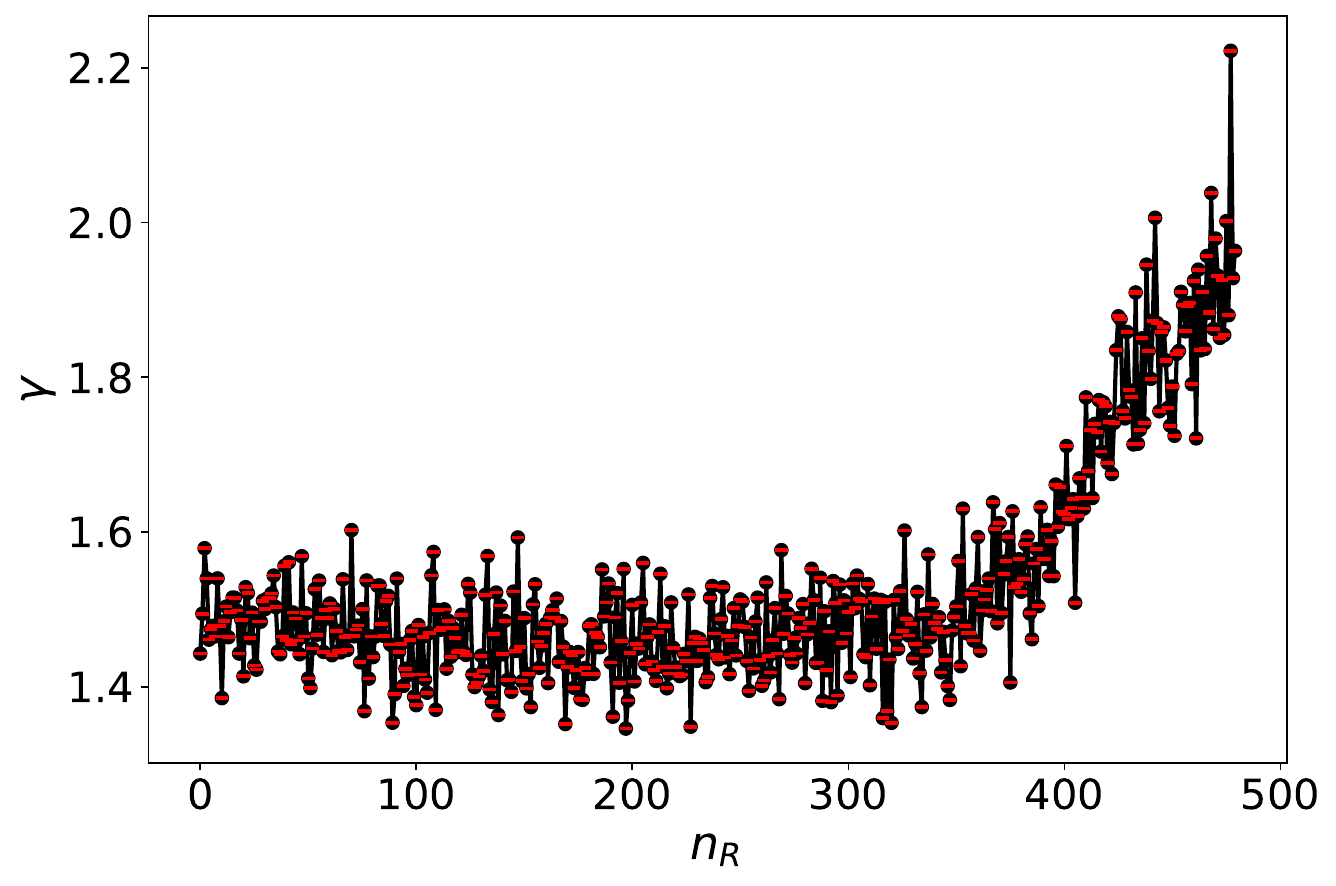}
			\caption{Critical exponents $\gamma$ for the PDF of dissipated
				energy in Fig.~\ref{pdf2d}, as a function of the number of
				rewirings $n_R$. The system consists of a
				$16\times16$ grid, with $n_R$ rewirings, and $10^5$
				iterations of the sandpile model after the critical state is
				reached. Then, results are
				averaged over 30 different sequences of $n_R$ rewirings.}
			\label{gamma_nr}
		\end{figure}

		In this case, after the stationary state is reached, the energy
		distributions for avalanches is a power-law, as seen in
		Fig.~\ref{pdf2d}, which is consistent with 
		Ref.~\cite{Bak}, and also holds for the rewired network. 
		The complementary CDF of the energy variations is shown in the Supplementary Material
		A linear fit of the curves yields a power-law with exponents
		$\gamma\sim1.45$, slightly increasing when the number of rewirings
		is large. Details are given in
		Fig~\ref{gamma_nr}.
		
\begin{figure}[H]
			\centering
			\includegraphics[width=\textwidth]{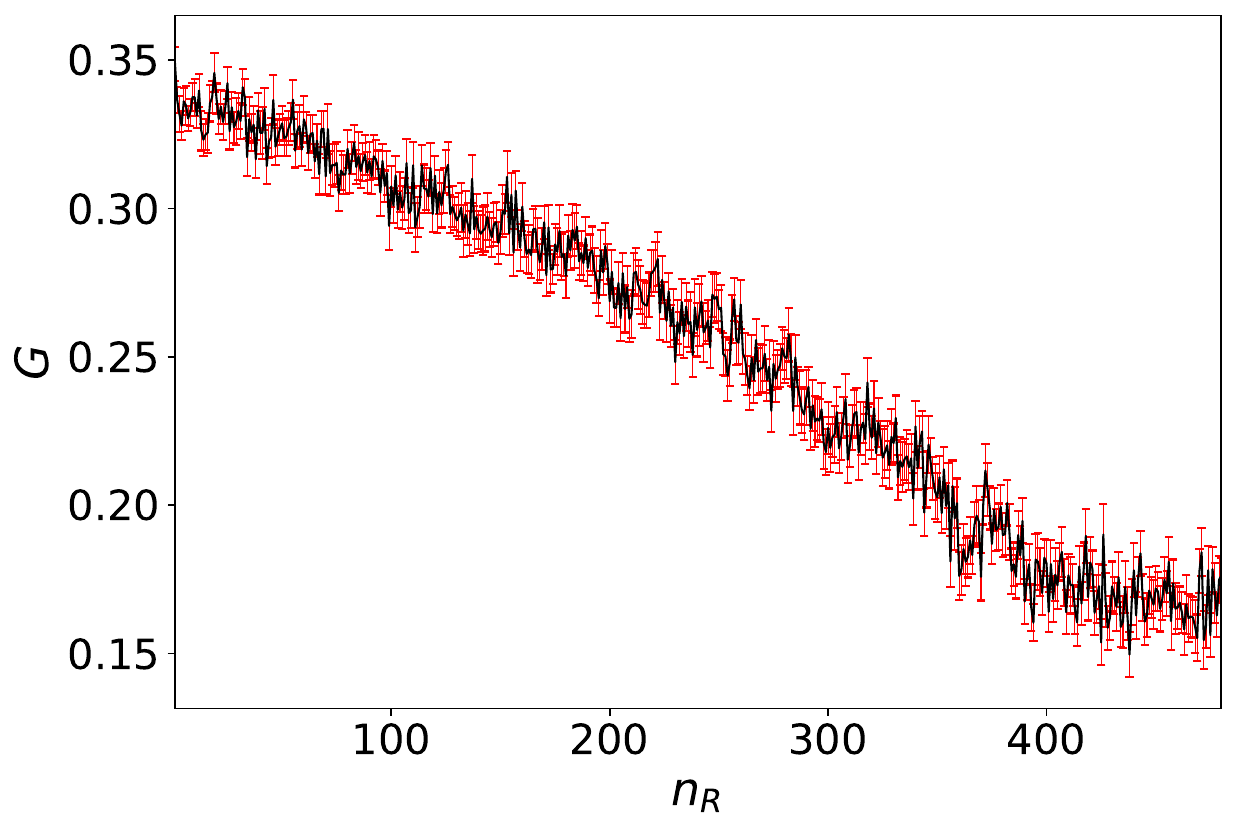}
			\caption{Gini coefficient for the distribution of energy in
				the critical state, as a function of the number of
				rewirings. The system consists of a 
				$16\times16$ grid, with $n_R$ rewirings, and $10^5$
				iterations of the sandpile model after the critical state is
				reached. Then, results are
				averaged over 30 different sequences of $n_R$ rewirings.} 
			\label{ginis2d}
		\end{figure}

		Unlike the 1D case, the Gini coefficient (see Fig.~\ref{ginis2d}) does
		not present a sharp 
		transition as the number of rewirings increases. Rather, it
		decreases until saturating at a minimum value after about 400
		rewirings. This occurs because, for such large values of the
		number of rewirings, nodes that remain to be rewired are
		located at the edges of the system, and those nodes have only one
		connection which can be modified. This leads to the change in behavior
		observed in Fig.~\ref{ginis2d}, as further modification of node
		inequality is slower.
		
\begin{figure}[H]
			\centering
			\includegraphics[width=\textwidth]{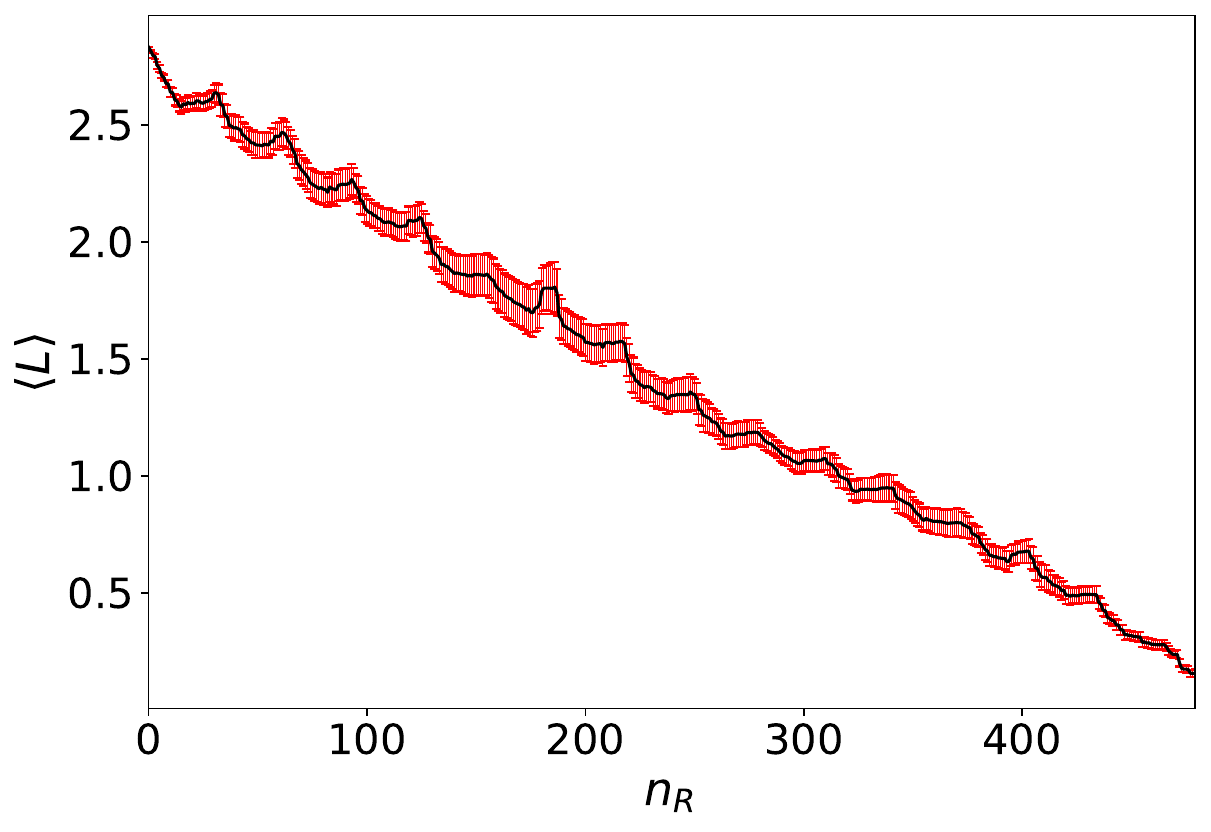}
			\caption{Average shortest path length of the network as a
				function of the number of rewirings. The system consists of a 
				$16\times16$ grid, with $n_R$ rewirings. Then, results are
				averaged over 30 different sequences of $n_R$ rewirings.}
			\label{meanpath}
		\end{figure}

		As in the 1D case, the shortest average
		path does not show a transition between two behaviors
		(Fig.~\ref{meanpath}), but now it decreases at a linear rate with the
		number of rewirings~$n_R$. The local maxima in
		Fig.~\ref{meanpath} are due to the way the network is updated,
		following the 
		labeling of the nodes, and because the change in average path length is
		different when rewiring involve nodes in the bulk or in the
		boundaries of the grid. In effect, since nodes at the boundaries must
		have one connection leading out of the system to ensure energy
		dissipation, rewiring does not have the same effect for such nodes, as
		compared with nodes in the bulk. This is seen as local maxima of the
		average path length every time one row of the grid is completely
		rewired.  Notice that, as mentioned in 
		Sec.~\ref{model}, our simulations have considered a square grid with
		16 nodes per side, which is consistent with the 16 maxima observed in
		Fig.~\ref{meanpath}. 
		\section{Size Effects}
		\label{finite_size}
		
		The results in the previous sections show a clear transition
		in the Gini coefficient for 1D, but not in 2D, whereas there is a
		transition for the critical exponent for 2D, but not 1D To better study this, and
		its possible dependence on size effects, we will study the evolution
		of the Gini coefficient as a funtion of
		$$f_r=n_r/N_c\ ,$$ 
		where $n_r$ is the 
		number of rewirings, and $N_c$ is the number of connections.
		
		\subsection{1D Case}
		
\begin{figure}[H]
			\centering
			\includegraphics[width=\linewidth]{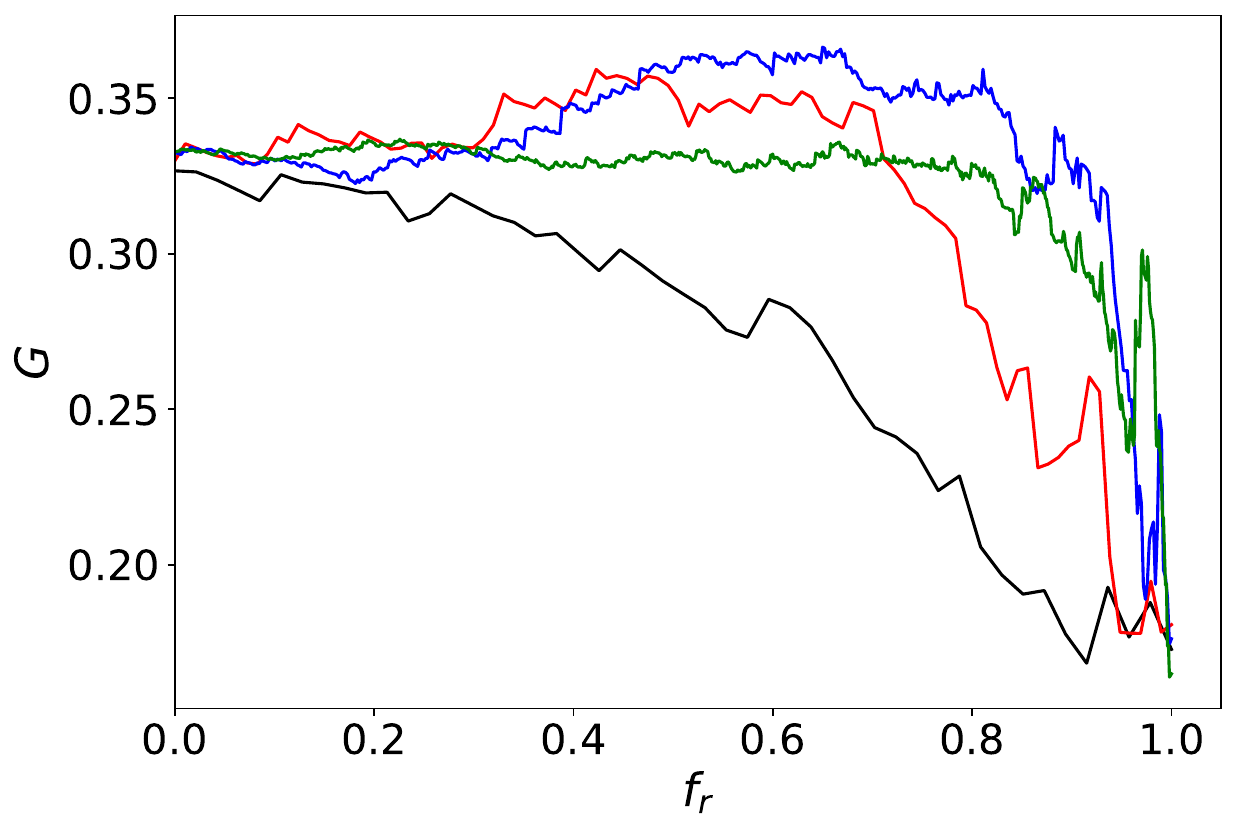}
			\caption{Gini coefficient for the distribution of energy in
				the critical state of the 1D model, as a function of the
				fraction of
				rewirings $f_r$. Each curve corresponds to a different system
				size: $N=50$ (black line), $N=100$ (red line), $N=500$ (blue
				line), and 
				$N=1000$ nodes (green line).} 
			\label{fig:giniorderparameter}
		\end{figure}

		In this case, we analyze the Gini coefficient for various network
		sizes, with $N=50,100,500,1000$ nodes, with $U=2$ and $Q=1$. We
		perform one simulation for each case, so curves are noisier than in the
		previous section, but a general trend can be appreciated. This is shown in
		Fig.~\ref{fig:giniorderparameter}.
		It is observed that, as the network size increases, the Gini
		coefficient tends to increase for all values of $f_r$, remaining
		approximately constant, until it collapses to a minimum value when
		all connections have been rewired. This implies that a sharper
		transition actually occurs for the Gini coefficient in the 1D case,
		but it is more evident for larger networks. The particular result
		previously shown in Fig.~\ref{gini} is consistent with this.

		We take the first derivative of $G$, $\left(dG/df_r\right)$, to
		determine where it has a maximum, which signals the critical value
		$f_v^c$ at which the transition occurs. To perform a robust numerical
		calculation of the derivative, given the stochastic fluctuations of
		the curves, it is necessary to reduce the noise prior to
		differentiation. For this, we apply a Savitzky-Golay filter, which
		smooths the data by fitting a polynomial to a subset of adjacent data
		points. In our approach, a third-degree polynomial is used in
		conjunction with a sliding, overlapping window of dynamic width. To
		preserve the sharpness of the transition, the window size is set to
		$1/6$ of the total data points for smaller datasets, up to a maximum
		fixed size of $21$ points for datasets larger than 126 points (always
		constrained to an odd integer, as required by the algorithm). 
		The results associated with the number of nodes and the control parameter at which the derivative is maximum are presented in the Table~\ref{tab:cutoff_fr2d}. 
		\begin{table}[H]
			\centering
			\caption{Critical value $f_r^c$ at which the Gini coefficient
				has maximum derivative, for various number of nodes $N$.}
			\label{tab:cutoff_fr2d}
			\begin{tabular}{|c |c|}
				\hline
				$N$ & $f_r^c$ \\
				\hline
				50   & 0.6809    \\
				100  & 0.8247    \\
				500  & 0.8471    \\
				1000 & 0.8766    \\
				\hline
			\end{tabular}
		\end{table}
		
		We find that the transition point depends on system size, but for
		large $N$  it stabilizes at around 85\% rewiring, which is consistent
		with Fig.~\ref{gini}. 
		
		To analyze this transition from a topological perspective, we examine
		some network metrics. In Sec.~\ref{results} it was already found that
		the  average shortest path length does not reveal a sharp signature of
		transition. Another metric related to shortest paths is the betweennes
		centrality $g(v)$, defined as
		$$
		g(v) = \sum_{s,t\in V}\dfrac{\sigma(s,t|v)}{\sigma(s,t)}\;,
		$$
		where $V$ is the set of nodes, $\sigma(s,t)$ is the number of shortest
		paths from node $s$ to node $t$, and $\sigma(s,t|v)$ is the number of
		those paths passing through a given node $v$ (other than $s$ and $t$).
		This metric quantifies the global influence of a node over the energy
		flow.
		
		Furthermore, we examine the in-degree ($\deg_{in}$) of the network.
		Because the total number of nodes and directed links is
		conserved during rewiring process, global averages such as the mean
		in-degree remain constant. Therefore, to capture how topological
		hetereogenity emerges, we focus on the variance and the maximum value
		of these structural metrics. This allows to quantify the
		inhomogeinity, and the possible concentration of incoming connections
		for some nodes.  
		
\begin{figure}[H]
			\centering
			\begin{minipage}{0.49\textwidth}
				\centering
				\includegraphics[width=\textwidth]{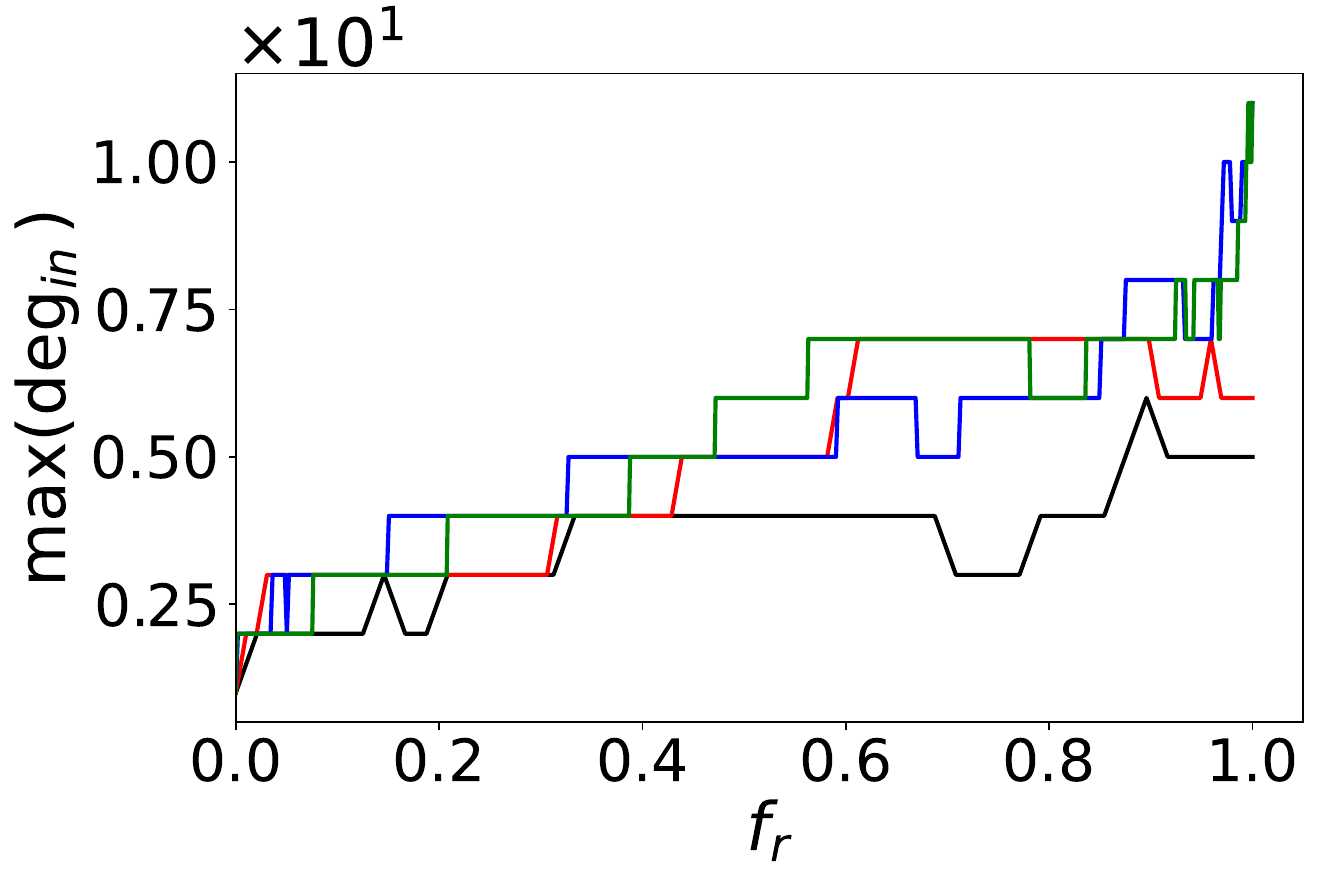}
				\label{fig:degreemax}
			\end{minipage}
			\hfill
			\begin{minipage}{0.49\textwidth}
				\centering
				\includegraphics[width=\textwidth]{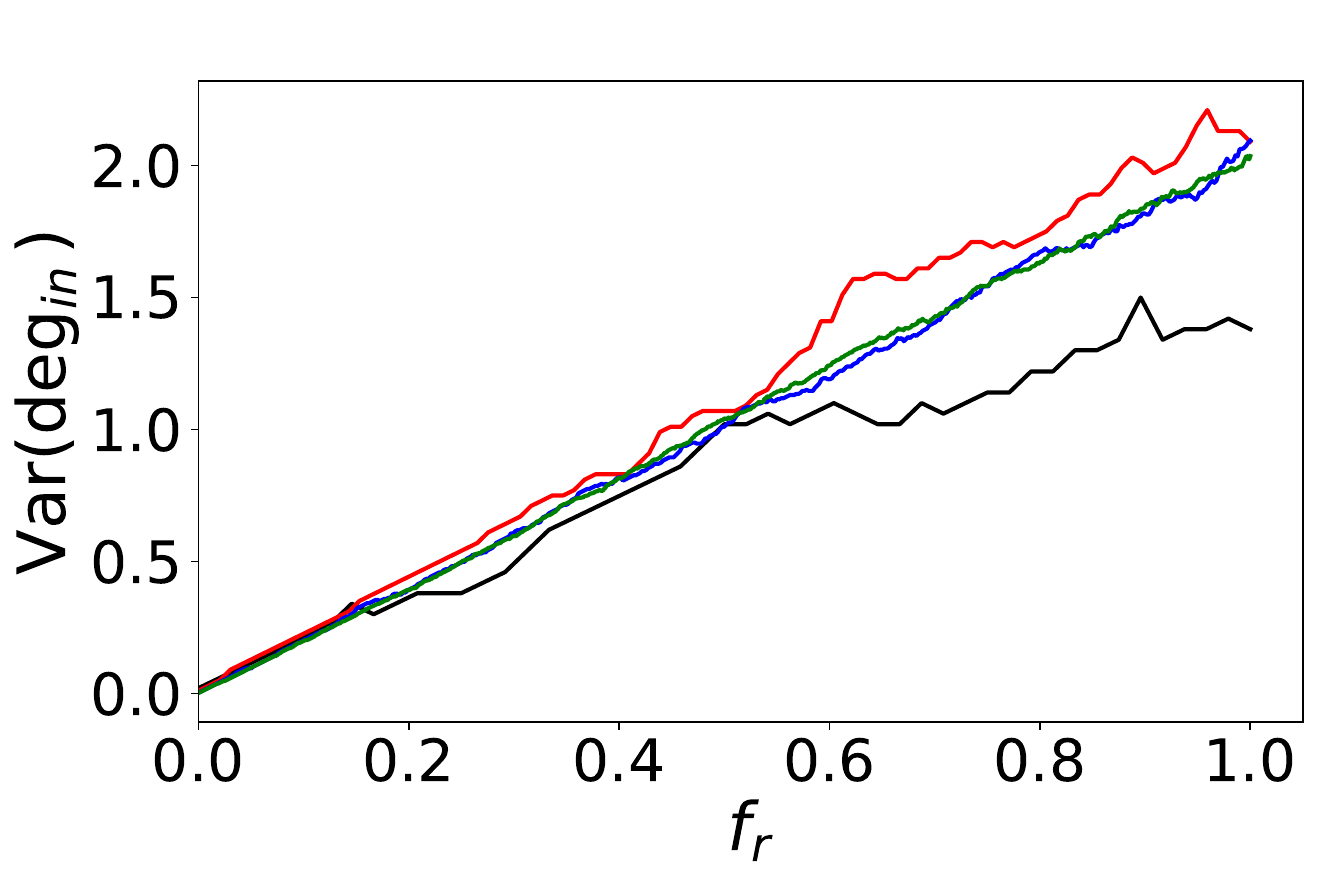}
				\label{fig:degreevar}
			\end{minipage}
			
			\vspace{0.5cm}
			
			\begin{minipage}{0.49\textwidth}
				\centering
				\includegraphics[width=\textwidth]{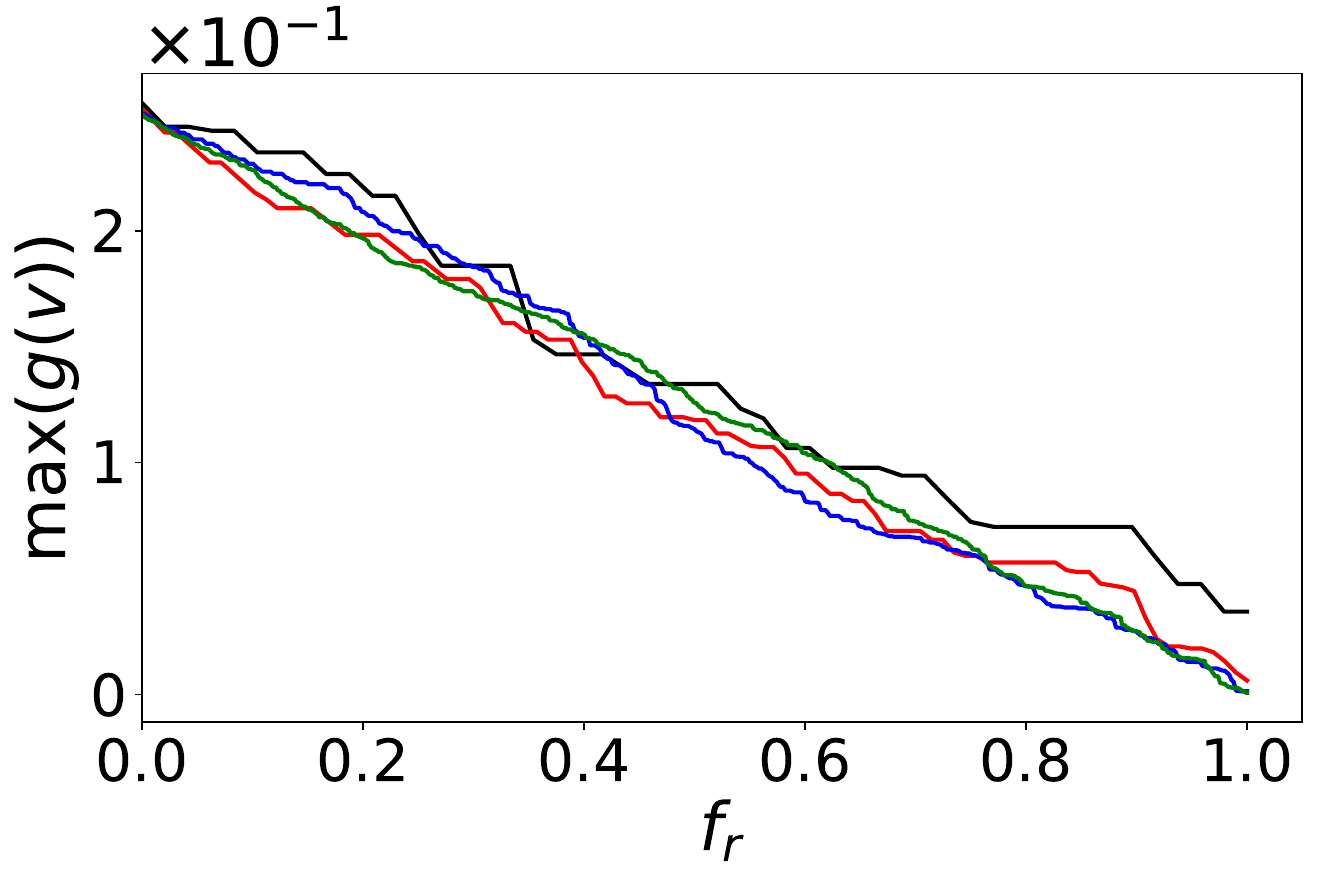}
				\label{fig:indegvariance}
			\end{minipage}
			\hfill
			\begin{minipage}{0.49\textwidth}
				\centering
				\includegraphics[width=\textwidth]{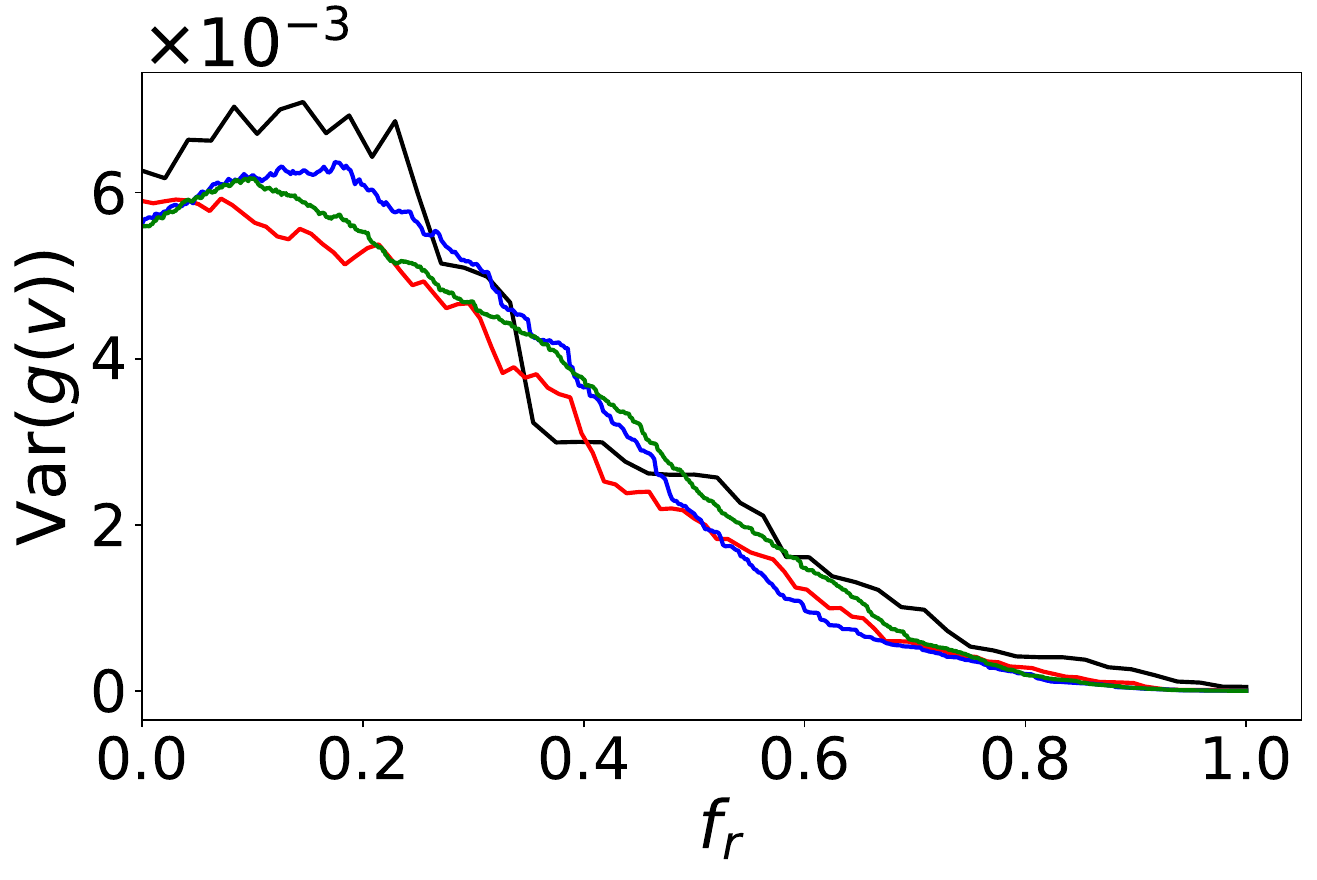}
				\label{fig:invariancebetweenness}
			\end{minipage}
			\caption{Maximum values and variance of the in-degree and the
				betweenness centrality, as a function of percentage of
				rewiring, for various network sizes. Top panel: in-degree;
				bottom panel: betweenness centrality; left panel: maximum
				value; right panel: variance. Network size: $N=50$ nodes
				(black line), $N=100$ (red line), $N=500$ (blue line), and
				$N=1000$ (green line).}
			\label{fig:combined}
		\end{figure}

		The evolution of these structural features is presented in
		Fig.~\ref{fig:combined}, which shows the maximum value and variance for
		both the in-degree (top panel) and betweenness centrality (bottom
		panel) as a function of the control parameter $f_r$. 
		
		From Fig.~\ref{fig:combined}, it is observed that both the betweenness
		centrality and its variance tend to zero as the rewiring increases.
		These means that there are essentially no prefered shortest paths
		beyond the critical point for the transition of the Gini coefficient,
		unlike the extreme case of the initial linear chain configuration,
		where there is 
		only one path for all energy releases. 
		
		As to the in-degree, figures show a slow increase of both the degree
		and the variance. Notice that degree can only have integer values,
		which explains the steps in the upper left panel. This behavior is 
		consistent with the network becoming less homogeneous as rewiring
		increases. 
		
		\subsection{2D Case}
		
		For two dimensions, Figs.~\ref{ginis2d} and~\ref{gamma_nr} showed a
		transition for the critical exponent, but not for the Gini
		coefficient. To consider size effects, we take networks with $N=16\times16$,
		$N=24\times24$, $N=32\times32$, and $N=40\times40$ nodes, with $U=3$
		and $Q=1$. As in the 1D case, one simulation per network was
		performed, but it is enough to see the general trend. 
		
\begin{figure}[H]
			\centering
			\includegraphics[width=\linewidth]{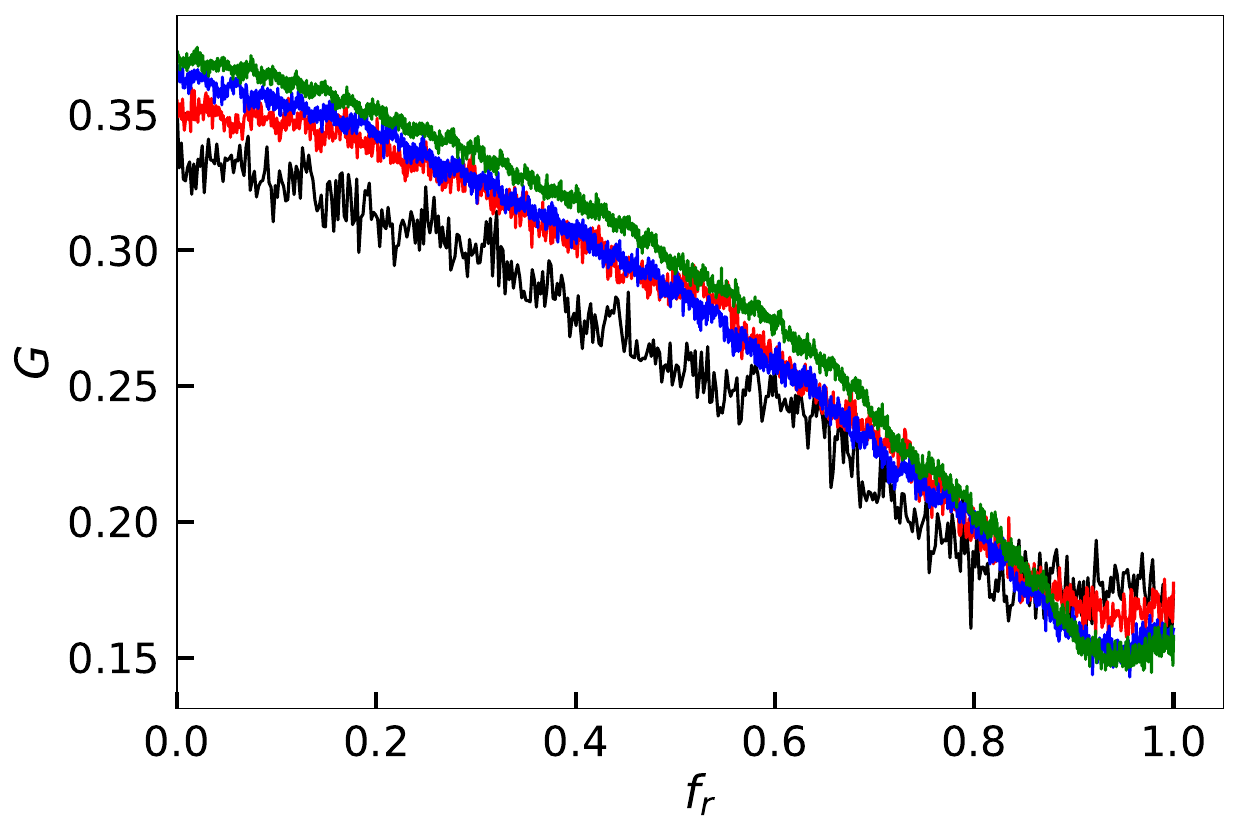}
			\caption{Gini coefficient as a function of the percentage of
				rewired links $f_r$, for different
				system sizes $N$: $N=16$ nodes (black line), $N=24$ (red
				line),  $N=32$ (blue line),
				and $N=40$ (green line).} 
			\label{fig:giniorderparameter_2D}
		\end{figure}

		The Gini coefficient for the grain distribution in the network is
		shown in Fig.~\ref{fig:giniorderparameter_2D}.
		At first glance, the Gini coefficient does not
		show the abrupt break that the 1D system has, regardless of network
		size. However, there is a change of
		behavior at large $f_r$. Smoothing the curves as in the 1D case, and
		calculating the maximum absolute value of the 
		derivative we find the corresponding critical value of the rewired
		proportion, $f_r^c$, at which this transition occurs. This is shown in
		Table~\ref{tab:cutoff_fr}. 
		\begin{table}[H]
			\centering
			\caption{Critical value $f_r^c$ at which the Gini coefficient
				has maximum derivative, for various number of nodes $N$, 2D case.}
			\label{tab:cutoff_fr}
			\begin{tabular}{|c |c|}
				\hline
				$N$ & $f_r^c$ \\
				\hline
				$16\times16$   & 0.7107    \\
				$24\times24$  & 0.7802    \\
				$32\times32$  & 0.8390    \\
				$40\times40$& 0.8771    \\
				\hline
			\end{tabular}
		\end{table}
		We notice a similar behavior to that observed for the 1D case, in
		Table~\ref{tab:cutoff_fr2d}. In effect, Table
		\ref{tab:cutoff_fr} also shows that the critical point has a weak
		dependence on system size, approaching $\sim$85\% rewiring for large
		networks. 
		The transition, 
		however, is different, as it goes from a monotonously decreasing
		behavior to an essentially constant, or slightly increasing, rather
		than from a constant behavior to a rapidly decreasing one
		(Fig.~\ref{fig:giniorderparameter}). It would be interesting to further
		explore the evolution of the transition point, for larger network
		sizes. However, based on our numerical experiments, one should expect
		that it should increase at a smaller rate, since
		$N$ is unbounded, whereas $f_r\leq 1$. 
		
\begin{figure}[H]
			\centering
			\begin{minipage}{0.49\textwidth}
				\centering
				\includegraphics[width=\textwidth]{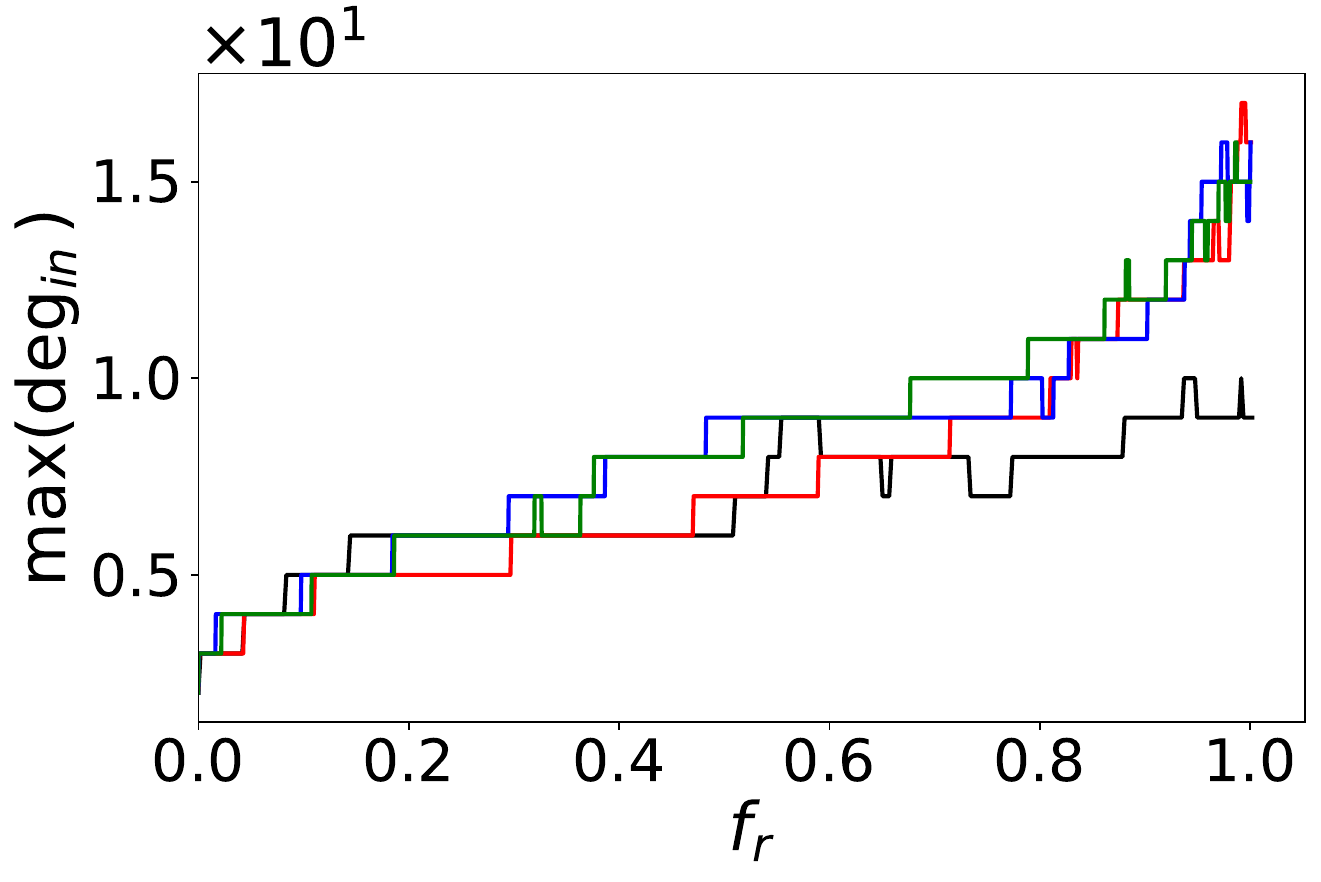}
				\label{fig:degreemax_2D}
			\end{minipage}
			\hfill
			\begin{minipage}{0.49\textwidth}
				\centering
				\includegraphics[width=\textwidth]{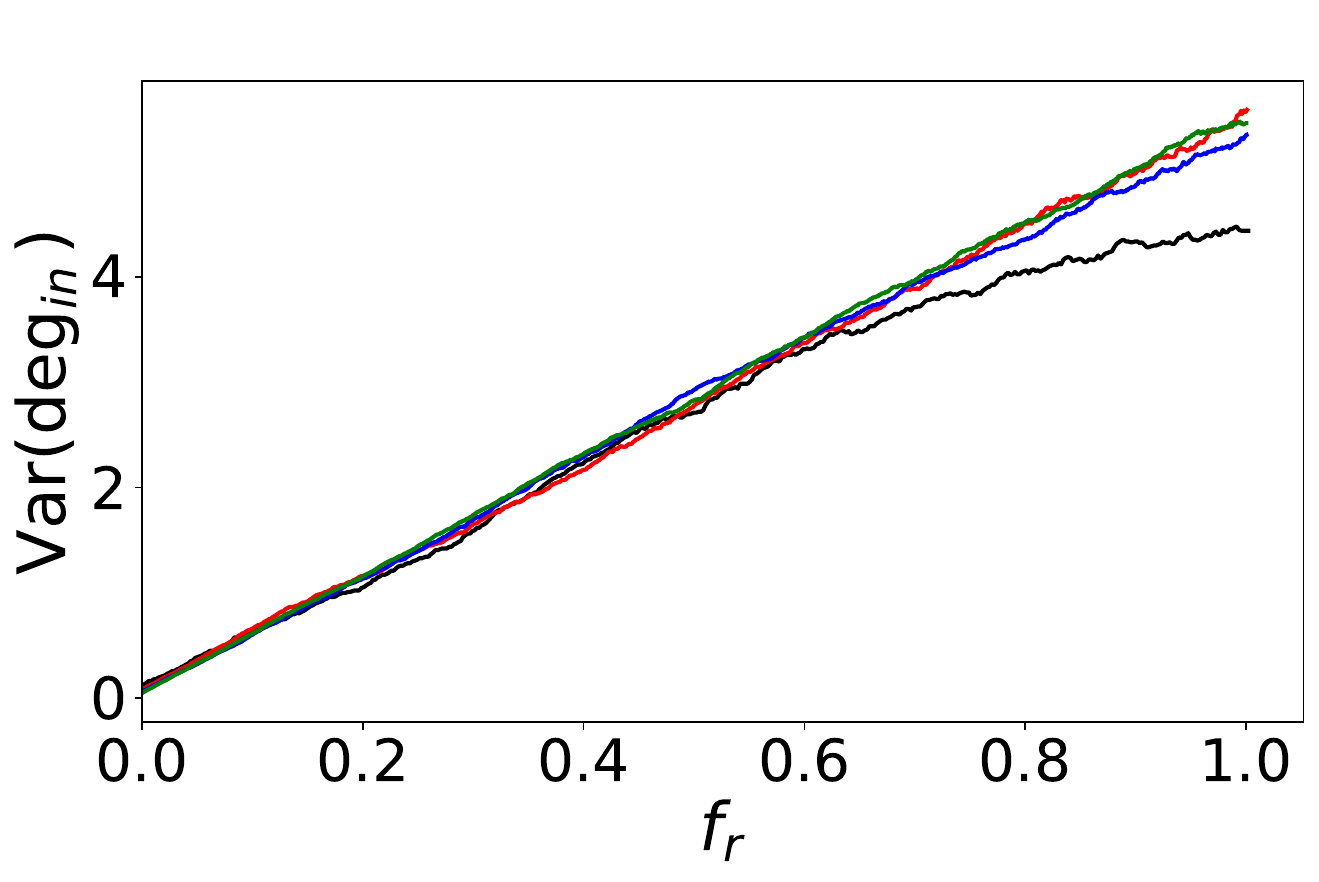}
				\label{fig:degreevar_2D}
			\end{minipage}
			
			\vspace{0.5cm}
			
			\begin{minipage}{0.49\textwidth}
				\centering
				\includegraphics[width=\textwidth]{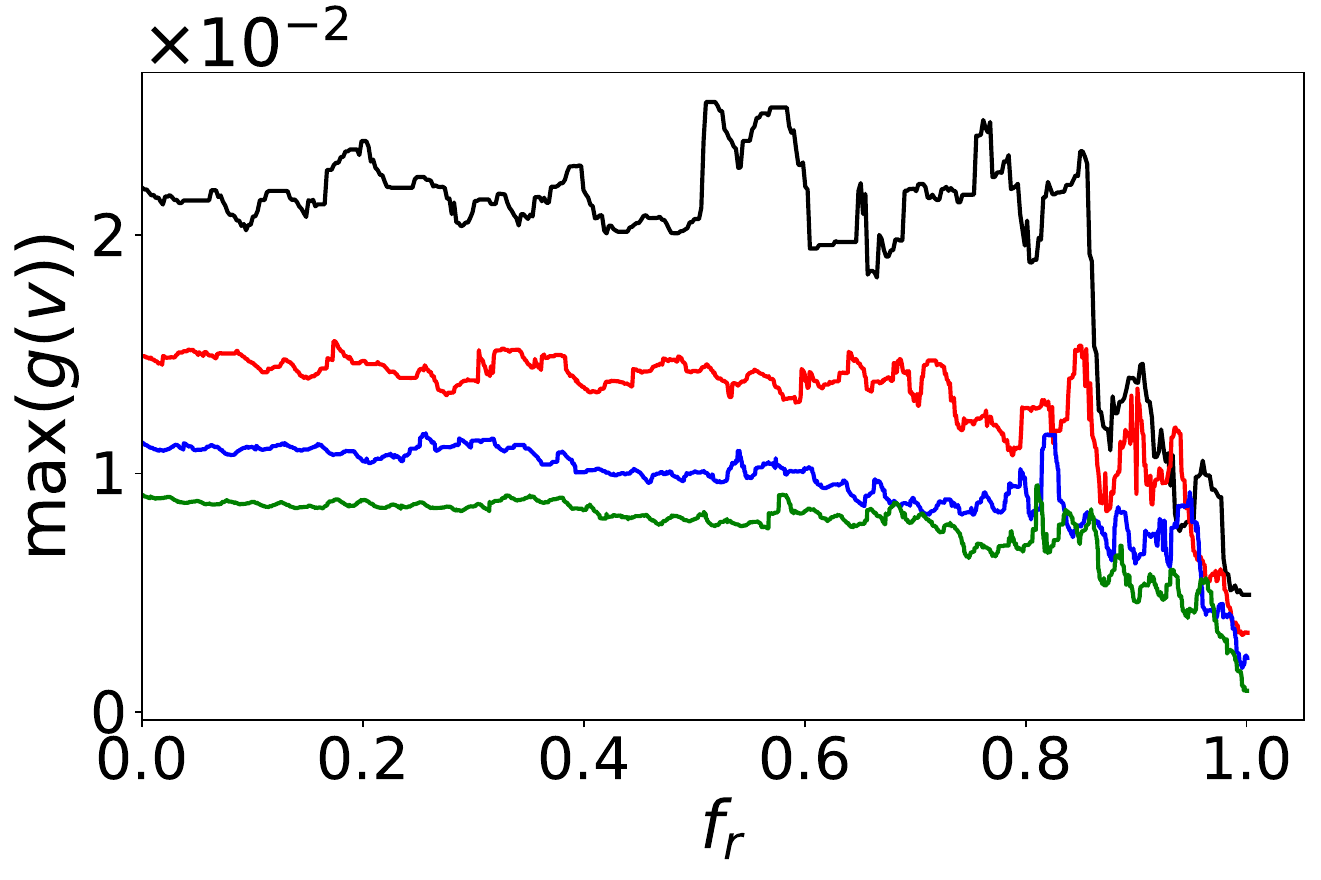}
				\label{fig:indegvariance_2D}
			\end{minipage}
			\hfill
			\begin{minipage}{0.49\textwidth}
				\centering
				\includegraphics[width=\textwidth]{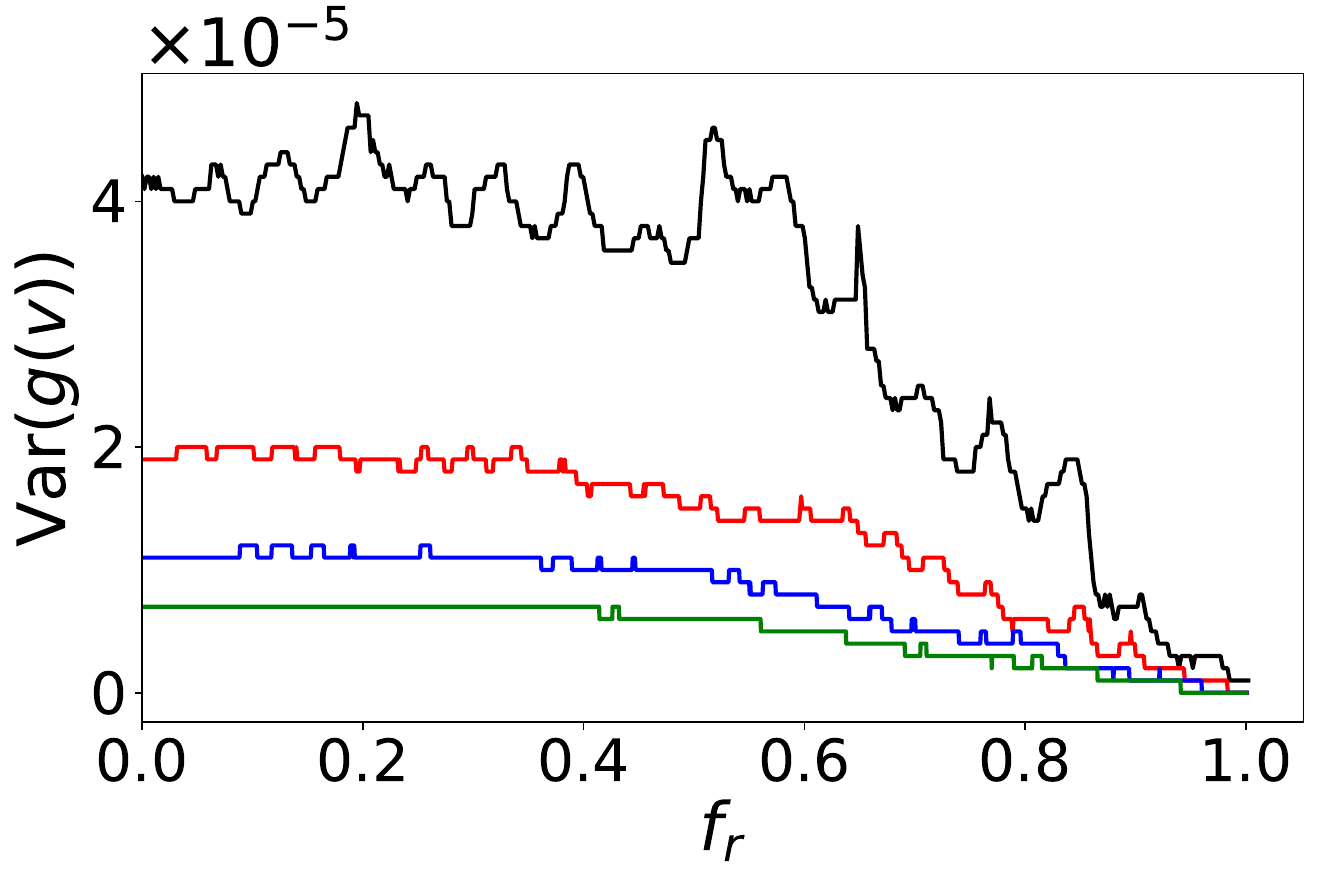}
				\label{fig:invariancebetweenness_2D}
			\end{minipage}
			\caption{Maximum values and variance of the in-degree and the
				betweenness centrality, as a function of percentage of
				rewiring, for various network sizes. Top panel: in-degree;
				bottom panel: betweenness centrality; left panel: maximum
				value; right panel: variance. Network size: $N=16\times 16$
				(black line), $N=24\times24$ (red line),
				$N=32\times32$ (blue line), and $N=40\times40$ (green line).}
			\label{fig:combined_2D}
		\end{figure}

		Figure~\ref{fig:combined_2D} is the analogous to
		Fig.~\ref{fig:combined}, showing the maximum value and the variance of
		the in-degree and the betweenness centrality. 

		The in-degree behavior is similar to the 1D
		case. 
		As to the betweenness centrality, both the maximum and the
		variance decrease, but after remaining essentially constant until,
		approximately, reaching the critical value for $f_r$. This result is
		consistent with that shown in Fig.~\ref{ginis2d}, where the Gini
		coefficient decreases monotonically until it reaches a point where it
		flattens into a nearly constant curve. 
		
		\section{Summary}
		\label{summary}
		
		In this work, we have studied, in a systematic way, the transition
		from the simple BTW sandpile to a sandpile over a directed network,
		where grains are added on the nodes, and grains are displaced to
		connected nodes following the directed edges. In general, choosing an arbitrary underlying network poses the risk of generating isolated nodes which accumulate loads instead of discharging them and closed topological loops. Both configurations trap energy, preventing the system from properly dissipating it even if the avalanche condition is locally satisfied. To address this issue, the model could be extended so that whenever a loop or an island emerges, one of the involved nodes is dynamically transformed into a sink. While this approach opens the door to more exotic network topologies, the present study focuses on a simple rewiring rule that inherently preserves the global degree and nodes and preserving SOC properties without the need to introduce new sinks. Thus, we randomly rewire nodes in such way that, at any stage of the network's evolution, energy release is always guaranteed once an avalanche is triggered. This contraint naturally leads to a tree-like structure
		with a single output node in 1D and with $2N-2$ output nodes in 2D,
		and where all nodes have a single discharge 
		path, while receiving discharges from several other nodes, depending
		on how the random rewirings occur.
		Then, we have studied various features of the
		resulting stationary state, and of the avalanche statistics, to
		determine how they are modified as the ``distance'' between the
		network and the BTW model increases, quantified by the number of
		rewirings performed.
		
		As more rewirings are introduced in the system, the number of
		discharge paths are increased, leading to shorter times to reach the
		stationary state. Also, since discharges can follow more than one
		path, the threshold condition to trigger the avalanches can be
		satisfied earlier than for the linear chain representing the BTW
		model, reducing the average load of nodes in the stationary state. 
		
		No power law or exponential behavior is found for the PDF of avalanche
		energy release events in 1D, whereas in 2D, a power law is found.
		These results are consistent with the original BTW model, and now we
		show that they persist
		even when the system is fully rewired.
		
		An interesting behavior is observed when the load
		distribution across the 1D network is characterized by the Gini
		coefficient. Then, a clear transition is observed at about 85\% the
		number of rewirings: for fewer rewirings, the Gini coefficient
		is essentially constant, and then, a sharp decrease is found. 
		As we show that by studying the behavior of the system as a function of
		the proportion of rewired nodes $f_r$, and changing the network size.
		Here, $f_r=0$ represents the initial, nonrewired system, and $f_r=1$
		is the case where all connections have been modified. We find that
		$dG/df_r$ has a maximum around  85\% rewiring. This is
		due to the easier distribution of load between nodes due to the larger
		number of discharge paths, but the transition is not evident by
		observing other features of the avalanche events, such as the PDF of
		energy releases, or metrics of the underlying network topology, such
		as the average distance between nodes, which decreases monotonically
		as the number of rewirings increases, without any particular
		transition. However, betweenness centrality, which is also related to
		shortest paths, but in a less trivial way, does indeed show a
		transition, where it is essentially null for all nodes, suggesting
		that there are no prefered nodes to connect two random nodes of the
		network. This occurs, as seen in Fig.~\ref{fig:combined},
		after 85\% rewiring, consistent with the transition observed for the
		load distribution.
		
		In the 2D case, a transition in the Gini coefficient
		occurs at about 400 rewirings in our case. It is interesting that this
		correlates with the
		behavior for the power-law exponent for the energy distributions,
		which remains at $\gamma\sim 1.45$, except after 400 rewirings,
		where it increases up to 2.21. This observation is further examined
		by the analysis of size effects, where it is shown that a smooth
		transition does indeed occur in the Gini coefficient, for different network sizes.
		Thus, the value of 400 found for the particular case studied in
		Sec.~\ref{results} is consistent with the critical value
		$f_r^c\simeq0.85$, as can be seen in Sec.~\ref{finite_size}. 
		
		We also notice that dimensionality has various
		consequences that can explain the fact that the transition is smoother
		for 2D with respect to the 1D case. As already observed in the 1D
		case, rewiring adds new discharge paths that decrease the energy of
		the stationary state, and the avalanche size. However, since the
		boundary is not changed, transport flows to the same final node. In
		the 2D case, rewiring has the same effect on stored and released
		energy, but more exit paths are available, since energy can be
		released at any boundary node. Thus, on average, modifying a
		particular connection should not
		have a strong effect on the distribution of load, and thus on the Gini
		coefficient due to the larger availability of discharge paths in the
		2D system.
		
		Our work suggests that the Gini coefficient of the grain distribution
		can be a useful way to examine the transition of transport properties
		in a complex network.  This metric, typically associated to the
		measurement of inequality in economics, is a useful way to characterize
		the distribution of a variable across a system by means of a single scalar
		value. In general, any scalar metric derived from the distribution function
		has less information, and thus the issue is whether the metric
		provides any useful insight. In our case, previous works had already shown
		the relevance of the Gini coefficient to yield information about the
		evolution of a network of sunstpots along the solar cycle, unlike other usual metrics
		such as the network 
		density or the clustering coefficient~\cite{Munoz_az}, and to reveal
		the transitions between the elastic and plastic regimes during
		deformations in metallic glass \cite{corvacho2024shear}, transitions which are not clear in the distribution
		functions themselves. 
		
		In this sense, it is interesting to observe that
		transitions in the Gini coefficient of relevant variables have also
		been observed in 
		other types of networks, such as in power grid networks where
		connections are removed as part of a robustness
		analysis of the network~\cite{wenli2016multi}, where the Gini
		coefficient shifts from a range of $G=0.8$ to $G=0.6$ to $G\approx
		0.32$. Other studies have considered systems more similar to ours, such as
		percolation on a square lattice, and the fiber bundle model of
		fractures~\cite{das2023critical}, also finding that the Gini
		coefficient reveals phase transitions in such systems.

		As to the shortest path, it decreases as in the 1D case, no transition
		is observed either, but the decrease is linear with the number of
		rewirings, except for sudden and local increases at specific
		values of $n_R$, which are related to the nodes
		where released from the system at the edges of the grid. These nodes have the fewest connections within the system since only one connection is directly connected to the network, while the rest of the nodes have two connections.
		
		Rewirings in this model affect the topology of the network and, in
		turn, its transport properties. These results could be relevant to
		study energy release along preferential paths, such as in magnetic
		reconnection events, thunderstorms, or money flow, processes where a
		certain variable is transferred between agents or 
		locations which are topologically, but not necessarily spatially close
		to each other.
		
		Finally, it is interesting to note that the exponents obtained for
		energy release are similar with some 
		obtained by models and observations of solar flares, which range
		between $\gamma = 1.44$ and $\gamma=2$ depending on the type of model
		and measurement technique~\cite{Aschwanden_b,Charbonneau,
			zamorano2025lu}; and in electrical discharges where the release of
		energy follows a power law with exponent $\gamma =
		2.3$~\cite{clark2021nonlocal}. These exponents for the two dimensional
		case, in addition to being consistent and falling within the range of
		sandpile models and these examples of physical systems, exhibit a
		saturation of approximately $85\%$, just like the two-dimensional Gini
		coefficient. This effect, both in the case of the Gini coefficient and
		in the case of the avalanche energy dissipation exponents, is
		explained by the fact that nodes closer to the last sink node are more
		likely to acquire new connections. Consequently, highly connected
		topological hubs emerge of these hubs while other become depleted. As
		$f_r$ increases, the in-degree of these hubs grows considerably, as
		shown in Fig.~\ref{fig:combined} and Fig.~\ref{fig:combined_2D}.
		
		This localized accumulation of connections has two fundamental
		dynamical effects. First, these hub nodes gain multiple energy input
		pathways, causing them to reach their critical threshold and
		subsequently release energy much more rapidly. Second, this rewiring
		alters the global energy transport, as evidenced by the betweenness
		centrality. As the rewiring fraction approaches the critical regime
		($f_r\sim f_r^c$), the variance of the betweenness centrality rapidly
		drops towards to zero. This vanishing variance alongside a decreasing
		maximum betweenness signals a topological homogenization where, on
		average, all nodes contribute equally to energy flow, leading to a
		more direct and efficient energy flow through 
		shorter escape paths to the system boundaries.  
		
		The present analysis could be useful for systems
		which both exhibit SOC features, and whose transport flow is directed
		on a nonhomogeneous network, as is the case of magnetic field lines
		directing flow from one magnetic active region to the opposite
		region, river basins from the mountains to the sea, or electrical
		discharges between clouds and the ground. Different network models
		and rewiring strategies should be necessary to study such systems, but
		previous work based on the Lu-Hamilton model for solar
		flares~\cite{zamorano2025lu} suggests that rewired networks pose 
		interesting issues for the study of complex systems exhibiting SOC
		features. 
		
		\section*{Acknowledgments}
		This project has been financially supported by FONDECyT under
		contract No.~1242013 (V.M.), and acknowledgment to ANID grant
		No.~21231335 (A.Z.). 
		\bibliographystyle{elsarticle-num} 
		\bibliography{sandpile_redes} 
		
		

	\end{document}